\documentclass[aps,prb,preprint,onecolumn,amsmath,amssymb,superscriptaddress,eqsecnum,floatfix,scrartcl]{revtex4-1}
\pdfoutput=1
\usepackage{amsmath}
\usepackage{amssymb}
\usepackage{graphicx}
\usepackage[tight]{subfigure}
\usepackage{enumitem}
\usepackage{bm}
\usepackage{todonotes}

\newcommand{\eg}{\textit{e.g.} }

\newcommand{\I}{\mathbb{I}}
 \DeclareMathOperator{\Tr}{Tr}

\newcommand{\norm}[1]{\left\lVert#1\right\rVert}
\renewcommand{\vec}[1]{\boldsymbol{\mathbf{#1}}}
\usepackage[font={footnotesize}]{caption}
\usepackage{hyperref}
\hypersetup{pdftex,colorlinks=true,allcolors=blue}
\usepackage{hypcap}

\graphicspath{{./}{./images/}}

\begin{document}

\title{Measuring the distance between quantum many-body wave functions}

\author{Xiao Chen}
\email{xchen@kitp.ucsb.edu} \affiliation{Kavli Institute for
Theoretical Physics, University of California at Santa Barbara, CA
93106, USA}

\author{Tianci Zhou}
\email{tzhou13@illinois.edu} \affiliation{Department of Physics
and Institute for Condensed Matter Theory, University of Illinois
at Urbana-Champaign, IL 61801-3080, USA}

\author{Cenke Xu}
\email{xucenke@physics.ucsb.edu} \affiliation{Department of
Physics, University of California at Santa Barbara, CA 93106, USA}

\date{\today}

\begin{abstract}

We study the distance of two wave functions under chaotic time
evolution. The two initial states are differed only by a local
perturbation. To be entitled ``chaos" the distance should have a
rapid growth afterwards. Instead of focusing on the entire wave
function, we measure the distance $d^2(t)$ by investigating the
difference of two reduced density matrices of the subsystem $A$
that is spatially separated from the local perturbation.
This distance $d^2(t)$ grows with time and eventually saturates to
a small constant.
%We use the operator scrambling picture to understand the growth of $d^2(t)$ in time and show that it describes the same physics as the square of commutator $C(t)$ (out-of-time-order correlator) in one dimensional system. Both $d^2(t)$ and $C(t)$ are proportional to the area of the wave front in subsystem $A$.
We interpret the distance growth in terms of operator scrambling
picture, which relates $d^2(t)$ to the square of commutator $C(t)$
(out-of-time-order correlator) and shows that both these
quantities measure the area of the operator wave front in
subsystem $A$.
%We further explicitly consider various spin chain models and find that for one dimensional spin-$\frac{1}{2}$ chain with non-local power-law interaction, $d^2(t)$ can have an appreciable {\it exponentially} increasing regime in time when the local perturbation and the subsystem $A$ are well separated.
Among various one-dimensional spin-$\frac{1}{2}$ models, we
numerically show that the models with non-local power-law
interaction can have an {\it exponentially growing} regime in
$d^2(t)$ when the local perturbation and subsystem $A$ are well
separated. This regime is absent in the
spin-$\frac{1}{2}$ chain with local interaction only. After
sufficiently long time evolution, $d^2(t)$ relaxes to a small
constant, which decays exponentially as we increase the system
size and is consistent with eigenstate thermalization hypothesis.
Based on these results, we demonstrate that $d^2(t)$ is a useful
quantity to characterize both quantum chaos and quantum
thermalization in many-body wave functions.

\end{abstract}

\maketitle

\section{Introduction and Motivation}
% quantum chaos, level repulsion

Quantum chaos is a subject that has attracted a lot of
attention and efforts from various subfields of physics. It is
important in understanding the quantum nature of black hole
dynamics and plays a key role in the process of thermalization in
a fully isolated quantum many-body system. The traditional method
of studying quantum chaos is to analyze the spectrum correlation
of the Hamiltonian. According to Bohigas-Giannoni-Schmidt
conjecture, the spectrum of the quantum chaotic model shows level
repulsion statistics which is universally determined by the random
matrix theory of the same symmetry
class \cite{BohigasGiannoniSchmidtConjecture}. However, the level
repulsion statistics is a static property of the Hamiltonian and
does not directly help us understand the dynamics of quantum
chaos.

%Quantum chaos has received a revival interest in many different areas. It characterizes the information scrambling in the black hole physics and is responsible for the thermalization of some fully isolated quantum many-body system. The traditional description of quantum chaos is through the level repulsion statistics of the Hamiltonian, which is conjectured to be universally determined by random matrix of the same symmetry class\cite{BohigasGiannoniSchmidtConjecture}. It is however a static property and not very informative in the quantum chaotic evolution.

In a classical chaotic system, the trajectories of two states with
a small initial separation can diverge exponentially under time
evolution. This phenomenon is called the butterfly effect and can
be characterized by the Lyapunov
exponent~\cite{Gutzwiller2013chaos}. In quantum mechanics,
physical states are wave functions in a Hilbert space. Then a
natural question to ask is how to define a ``distance" between the
wave functions, and use the distance to characterize quantum
chaotic systems. A sensible definition of distance should
reproduce the diverging behaviors under a quantum chaotic
evolution.

The goal of this work is to give a proper definition of the
notion of distance between wave functions to characterize quantum
chaos. More specifically, we prepare two identical wavefunctions
and perturb one of them with a local operator $\hat{O}(x)$, then
evolve them with the same chaotic Hamiltonian, then we expect
a proper definition of ``distance" will grow with time. Of
course, one simple definition of ``distance" is the overlap
between the wave functions. However due to the unitarity of
quantum evolution, the wavefunction overlap is time independent
and hence can not display any growth under time evolution,
including chaotic evolution. It is thus not a useful quantity to
characterize quantum chaos.

On the other hand, the
evolution of the reduced density matrix $\hat{\rho}_A$ is not
unitary.
%Therefore various distances of the reduced density matrices of the two wavefunctions will be dependent.
% In this paper,
We therefore measure the Hilbert-Schmidt distance $d^2(t)$ of the
two reduced density matrices and use this to quantify the
difference of two states. Our approach is based on the following observation: if we
choose a subsystem $A$ that is spatially separated from the local
operator $\hat{O}(x)$, then it will take some time for a local
observer in $A$ to ``feel" the difference between the two initial
states. 
%Thus we choose the reduced density matrix of the subsystem
$A$ as the starting point of our study.  
The same quantity was first proposed in
Ref.~\onlinecite{Leviatan_2017}, where the authors  used $d^2(t)$ to study the chaos dynamics  in some one dimensional spin-$1/2$ chain model with conserved energy. In this paper, we will continue to explore this quantity in more detail, discuss its physical interpretation in the language of operator spreading and also make connection with the square of commutator (out-of-time-order correlator)\cite{Larkin1969}.

To demonstrate this idea, we compute the time dependence of
$d^2(t)$ and observe the growth pattern and saturation value in
spin-$\frac{1}{2}$ one-dimensional chains with a finite length.
Not surprisingly, the scaling behavior of $d^2(t)$ is model
dependent. In spin models with non-local interactions\cite{Hastings2006,Gong2014}, we observe
a clear exponential growing regime after the perturbation has
propagated to the subsystem. By contrast such exponential
regime is absent in spin models with only local interactions. This
is slightly different from classical chaos, where the separation
of trajectories will always grow exponentially with time. In both
cases, we find that the saturation values are the distance of two
independent random pure states (Page states)\cite{Page1993},
meaning that the initial similarities of the two states have been
almost completely washed out after a long time chaotic
evolution.

The growth behaviors of $d^2(t)$ can be understood in the operator
spreading picture of the time evolved perturbation operator
$\hat{O}(x, -t)$ developed in Refs.~\onlinecite{Keyserlingk2017}
and \onlinecite{Nahum2017}. The operator $\hat{O}(x,-t)$ comprises
of operator basis that is spreading under the chaotic evolution.
Using the tools in quantum information, we are able to relate the
average distance $\overline{d^2(t)}$ over the initial Page states
to the area of the wave front of the operator basis in $O(x,-t)$
in subsystem $A$. The interaction controls the shape of the wave
front, and hence the growth pattern of $\overline{d^2(t)}$.

Based on this picture, we show that $\overline{d^2(t)}$ is related
to the square of commutator\cite{Larkin1969}
\begin{equation}
C(t)=-\Tr\left\{\left[ \hat{O}(x,t), \hat{O}^\prime(y) \right]^2\right\}
\end{equation}
which also measures the area of the wave front for one dimensional
chaotic models. This quantity can be used to detect quantum chaos
and is shown to grow exponentially with time in some large $N$
systems\cite{Shenker2013a,Shenker2013b,Shenker2014,Maldacena2016,Kitaev2014,Kitaev2015,Sachdev1993}.
The growth rate is the quantum analog of the Lyapunov exponent and
characterizes the quantum butterfly effect. Our argument relating
$C(t)$ with $\overline{d^2(t)}$ is supported by the numerical
collapse of the two curves (after rescaling) in the models we
study. Therefore $d^2(t)$ is a legitimate candidate to quantify
quantum chaos.

Although the growth of $d^2(t)$ is model dependent, after
sufficient long time evolution, once the wave front of
$\hat{O}(x,t)$ fully spreads into region $A$, $d^2(t)$ saturates
to a small constant, which decreases exponentially with the total
system size. We further study $d^2(t)$ for two different initial
states with the same energy in a model with static Hamiltonian. We
find that $d^2(t)$ always relaxes to an exponentially small
constant at late time. These results are consistent with
eigenstate thermalization hypothesis
\cite{Srednicki1994,Deutsch1991}, which states that for a generic
chaotic system, the reduced density matrix for a small subsystem
can eventually approach a thermal form with the effective
temperature set by the initial energy of the state.

%\todo[inline]{cite SYK paper}
%Admittedly, the distance in our paper is not the only candidate. The quantum computational complexity\cite{Susskind_2014, Brown2017} is also a distance measure of the wavefuntions.
%%It is defined to be the minimal number of unitary quantum gates selected from a universal computing set to transform one state to the other.
%For any two states $|\psi_1\rangle$ and $|\psi_2\rangle$, they can be connected by some unitary operator $\hat{U}$. Quantum complexity computes the minimal number of the unitary two-site gates to generate the $\hat{U}$ operator. This quantity is  interesting and can be mapped to geodesic motion in hyperbolic space. In the holographic systems, there are proposals like complexity-action duality, that enables the computation of complexity in terms of the bulk gravity action. %The fact that black hole saturates the complexity increasing bound is also a profound result.
%However, unlike $\overline{d^2(t)}$, the computation of complexity in the discrete spin systems is still hard to be performed and depends on the choice of the sets of gates.

%We expect that in the future this practical difficulty may be overcome by a continuum description in terms of the geodesic length in hyperbolic geometry \cite{Brown2017}.
%\todo[inline]{black hole saturates the bound}

%\todo[inline]{xiao: although $\hat{O}(x,t)$ relates the two states in our case, but the unitary operator is not unique. So the gates do not necessarily generate $\hat{O}(x,t)$. }

We first briefly summarize the main results of this paper: (1)
$d^2(t)$ measures the distance between the wave functions and has
similar scaling behavior as the square of commutator. It can grow
exponentially in time in spin-$\frac{1}{2}$ chain models with
non-local power-law interactions. (2) The saturation value of
$d^2(t\to\infty)$ is a small constant which characterizes the
quantum thermalization in many-body systems.

This paper is structured as follows. We first define the
Hilbert-Schmidt distance $d^2(t)$ between two wave functions in
Sec.~\ref{definition} and then numerically study $d^2(t)$ in two
spin-$\frac{1}{2}$ chain models with local interactions in
Sec.~\ref{local_interaction}. In Sec.~\ref{power_law_interaction},
we further study $d^2(t)$ in spin models with non-local power-law
interaction. In Sec.~\ref{op_scramble}, we give a physical
interpretation for $d^2(t)$ in terms of the operator scrambling
picture and make the connection with the square of commutator. We
summarize and conclude in Sec.~\ref{conclusion}. The appendices
are devoted to the details of the calculations and techniques used
in this paper.

%%% Local Variables:
%%% TeX-master: "d_squ"
%%% TeX-PDF-mode: t
%%% End:

%We further explicitly consider various spin chain models and find that for one dimensional spin-$\frac{1}{2}$ chain with non-local power-law interaction, $d^2(t)$ can have an appreciable {\it exponentially} increasing regime in time when the local perturbation and the subsystem $A$ are well separated.

\section{Definition of the Hilbert-Schmidt distance}
\label{definition}

As we explained in the introduction, we prepare two initial
states $|\psi_1\rangle$ and
$|\psi_2\rangle=\hat{O}(x)|\psi_1\rangle$, where $\hat{O}(x)$ is a
local perturbation at position $x$. Our goal is to define a
distance between these two states during the evolution under the
same unitary operator $\hat{U}(t)$.

Intuitively, the initial distance should be small because
these two states only have a local difference. Under unitary time
evolution, we can write $|\psi_2 (t)\rangle $ as $\hat{O}(x, -t) |
\psi_1 (t) \rangle $, where $\hat{O}(x, -t ) =  \hat{U}(t)
\hat{O}( x) \hat{U}^{\dag}(t)$ is the backward evolved Heisenberg
operator. This operator becomes more and more non-local as
evolution, suggesting a growing distance between the states. A
properly defined distance should reflect these properties.

A na\"ive trial is to use the norm of the difference
\begin{equation}
d^2 \sim \norm{ |\psi_1(t)\rangle  - |\psi_2(t) \rangle }_2^2
\end{equation}
where $ |\psi_i(t)\rangle= \hat{U}(t) |\psi_i \rangle$ ($i = 1,
2$) are time evolved states. The 2-norm is related to the overlap
of $\langle \psi_2(t) | \psi_1(t) \rangle $, from which we can see
that the distance does not satisfy either of the requirements
proposed above. First we notice that the initial states can be
orthogonal even if $\hat{O}(x)$ is only a 1-bit operation. This
gives zero overlap and large initial separation. More importantly,
the overlap does not change under the {\it unitary} time evolution
and we end up with a constant distance.

These problems will not occur if we only pay attention to part of
the wavefunction -- the subsystem. A local perturbation outside
the subsystem will give an identical initial reduced density
matrix, thus giving a zero initial distance. Furthermore we note
that the dynamics of the subsystem is not unitary: the time
dependent {\it reduced} density matrices $\hat{\rho}_1(t) $,
$\hat{\rho}_2(t)$ of  $|\psi_1(t)\rangle$, $|\psi_2(t)\rangle$ do
not obey the Heisenberg equation. This motivates us to use
Hilbert-Schmidt distance (square-norm distance) between these two
reduced density matrices
\begin{align}
d^2(\hat{\rho}_1,\hat{\rho}_2)\equiv
\mbox{Tr}(\hat{\rho}_1(t)-\hat{\rho}_2(t))^2
\end{align}
as a measure of separation of states. This quantity is
semi-positive definite and satisfies the triangle inequality,
thus it is a well-defined distance. We will use it to explore the
quantum chaos dynamics.

\begin{figure}%[hbt]
\centering
\includegraphics[width=.62\textwidth]{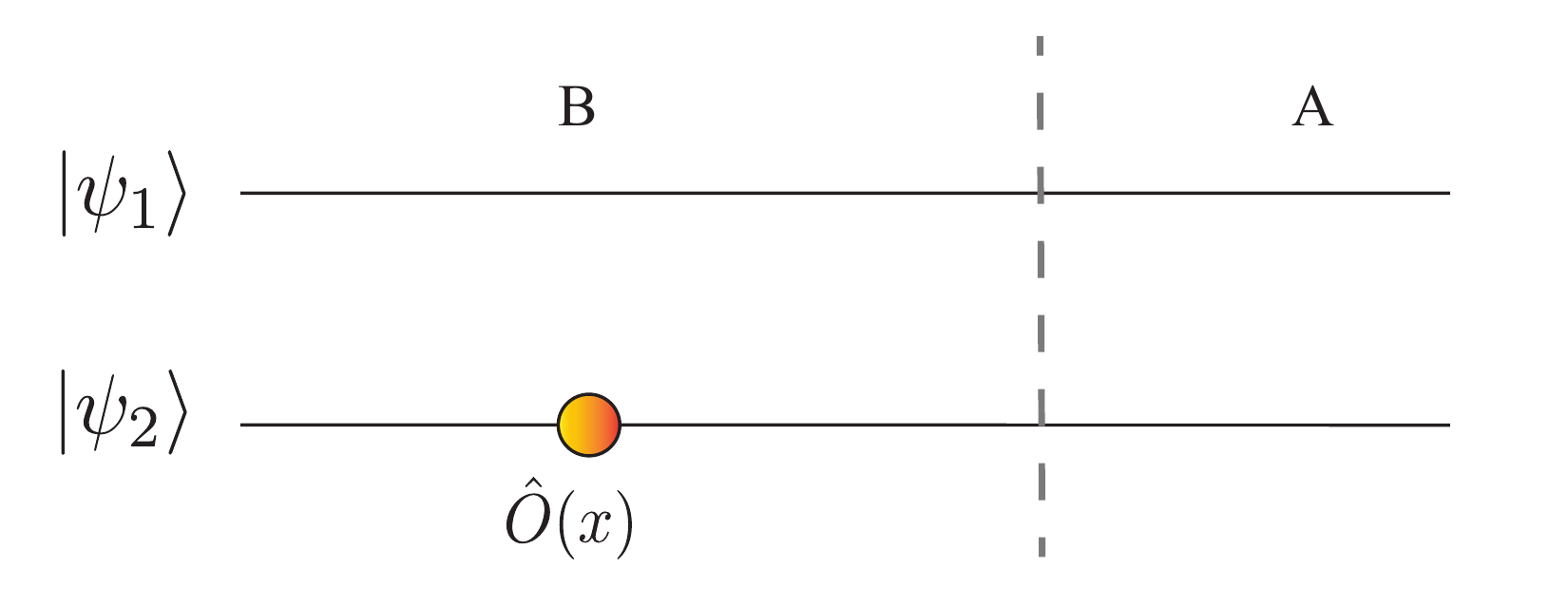}
\caption{The setup for computing $d^2(t)$.}
\label{fig:schematic}
\end{figure}

We expect the following development of $d^2(t)$ for the one
dimensional system shown in Fig.~\ref{fig:schematic}. The local
perturbation $\hat{O}(x,t=0)$ is outside of subsystem A, so at
$t=0$, the distance $d^2(t=0)$ is strictly equal to zero. As time
evolves, the operator $\hat{O}(x,-t)$ spreads out in the
space and as it reaches the subsystem $A$, $d^2(t)$ becomes
nonzero and increases with time. After sufficient long time,
$\hat{O}(x,-t)$ has spread over the entire space and $d^2(t)$ will
saturate to a constant.

%%%%%%%%%%%%%%
%%%%%%%%%%%%%%

%We further explicitly consider various spin chain models and find that for one dimensional spin-$\frac{1}{2}$ chain with non-local power-law interaction, $d^2(t)$ can have an appreciable {\it exponentially} increasing regime in time when the local perturbation and the subsystem $A$ are well separated.

\section{The spin chain models with local interaction}
\label{local_interaction}

\subsection{Floquet spin-$\frac{1}{2}$ chain}

The dynamics of the Floquet (periodically driven) system is
determined by the unitary time evolution operator over one period,
namely the Floquet operator. Following Ref.~\onlinecite{Kim_2014},
we consider the following Floquet operator:
\begin{align}
\hat{U}_F=\exp[-i\tau \hat{H}_z]\exp\left[-i\tau\hat H_x\right] ~,
\label{flo_op}
\end{align}
where
\begin{align}
&\hat{H}_x=\sum_{j=1}^{L}g\hat\sigma_j^x\nonumber\\
&\hat{H}_z=\sum_{j=1}^{L-1}\hat\sigma_j^z\hat\sigma_{j+1}^z+\sum_{j=1}^Lh\hat\sigma_j^z
~.
\end{align}
This model is a one dimensional periodically driven system with
period $T=2\tau$. In the numerical calculation, we choose open
boundary condition with number of sites $L=24$. The system
parameters are $(g,h)=(0.9045, 0.8090)$. The parameter $\tau$
controls the period of the Floquet operator.

For both numerical and analytical convenience, we choose the initial state
$|\psi_1\rangle$ as the Page state, which is defined as,
\begin{align}
|\Psi(\{\alpha_i \})\rangle=\sum_i \alpha_i|C_i\rangle
\end{align}
where the coefficients $\alpha_i$ of the state in a fixed basis
$\{ |C_i\rangle \}_i$ are random complex numbers subject to the
normalization constraint, with a probability distribution
invariant under a unitary basis transformation. This state has
(almost) maximal entanglement entropy and was first studied in
detail by D. Page\cite{Page1993}. $|\psi_2\rangle$ is obtained by
acting a local Hermitian operator $\hat{O}(x)=\hat{\sigma}^x(x)$
at location $x$ on $|\psi_1\rangle$. We study the Hilbert-Schmidt
distance $\overline{d^2(t)}$ between $|\psi_1(t)\rangle$ and
$|\psi_2(t)\rangle$ under the unitary time evolution, where the
over-line denotes averaging over the ensemble of the initial
Page states $|\psi_1\rangle$.

\begin{figure}[hbt]
\centering
 \subfigure[]{\label{fig:Floq_loc_d2_tau} \includegraphics[width=.32\textwidth]{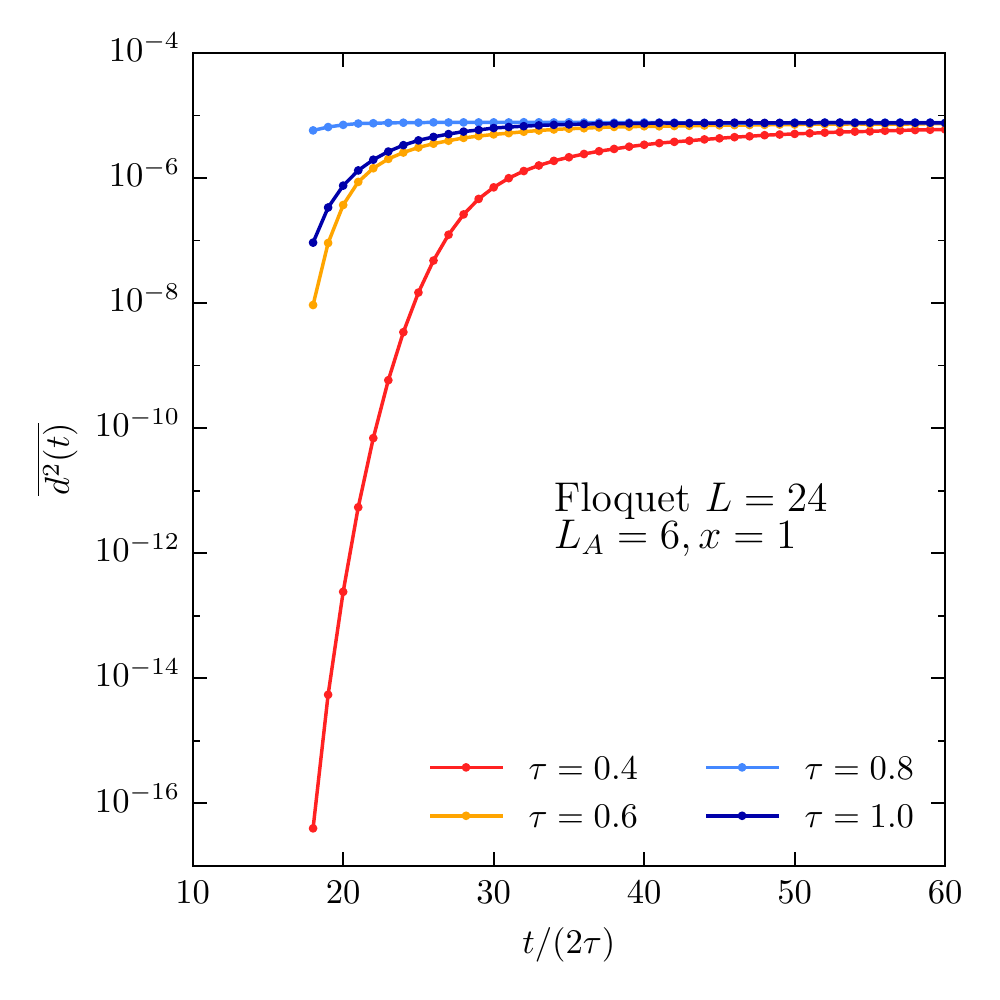}}
 \subfigure[]{\label{fig:Floq_loc_d2_x} \includegraphics[width=.32\textwidth]{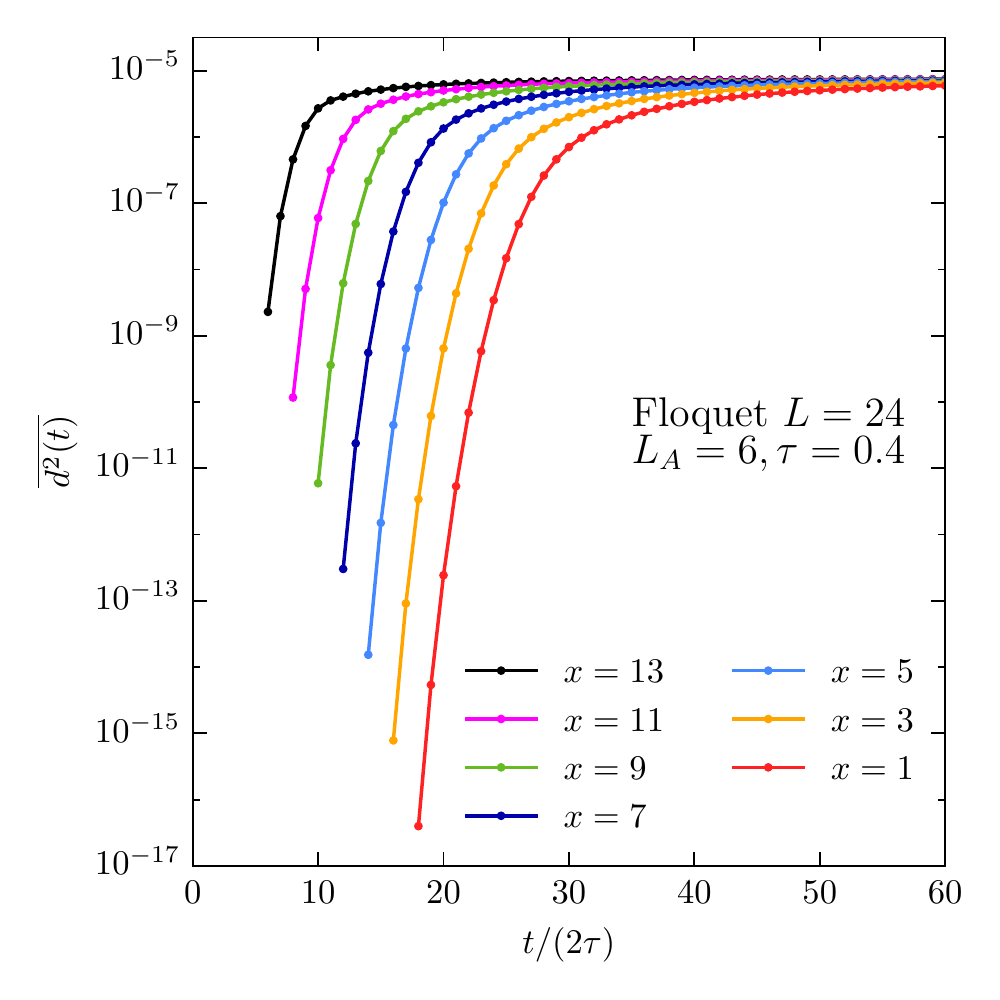}}
 \subfigure[]{\label{fig:Floq_loc_com} \includegraphics[width=.32\textwidth]{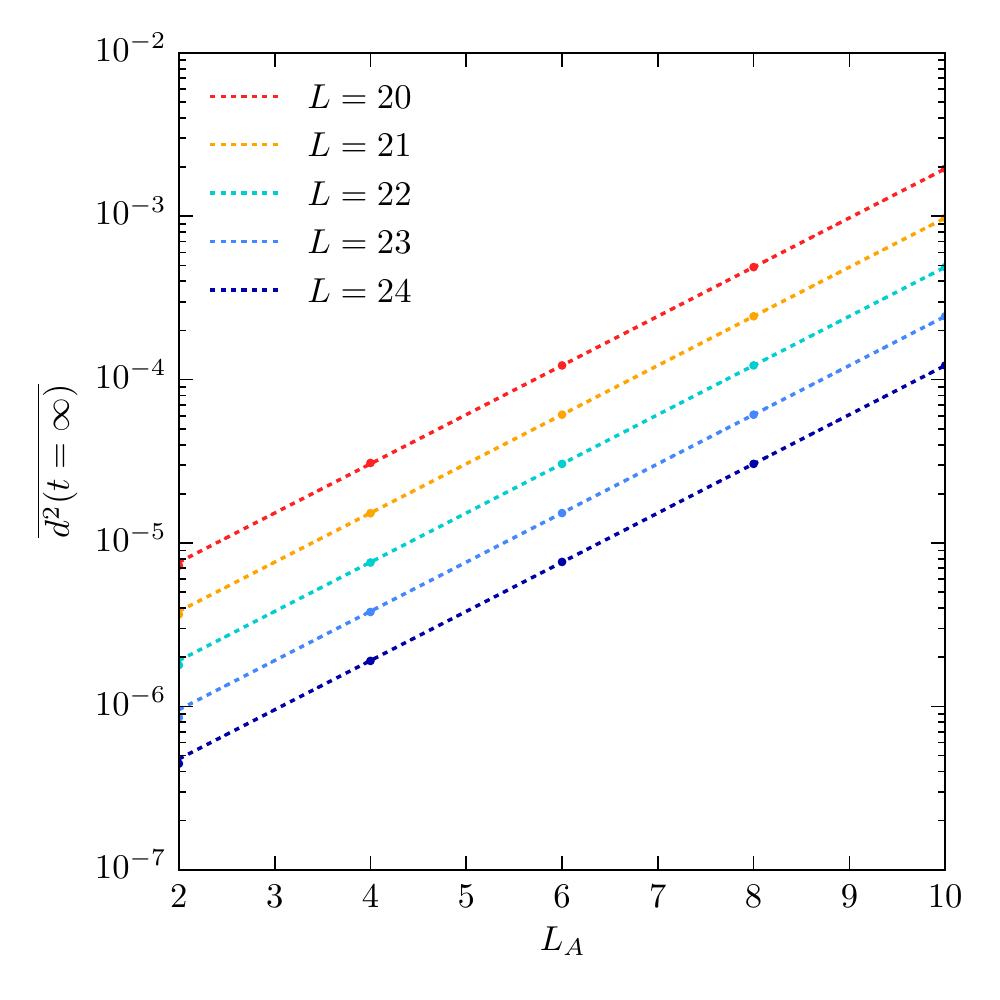}}
\caption{ (a) $\overline{ d^2(t)}$ vs the time step $t/(2\tau)$
for various $\tau$ with fixed $L_A$ and $x$. (b)
$\overline{d^2(t)}$ vs $t/(2\tau)$ for various $x$ with fixed
$L_A$ and $\tau$. (c) Comparision between the saturation value of
$\overline{ d^2(t)}$ and $\overline{d^2(\hat\rho_1,\hat\rho_2)}$
between two independent Page states in
Eq.\eqref{eq:analy_sat_val}. } \label{fig:Floq_local}
\end{figure}

Since we are considering a Floquet model, the evolution time $t$
is an integer multiple of period, $t=nT$ with $n\in \mathbb{Z}_+$.
There are in total three parameters we can tune: the half period
of Floquet system $\tau$, the length $L_A$ of the subsystem A and
the location of the local operator $x$. We first fix $L_A=6$,
$x=1$ and change $\tau$. As shown in
Fig.~\ref{fig:Floq_loc_d2_tau}, when $t/ T <18$, $\hat{O}(x,-t)$
does not propagate to region $A$ and we have $\overline{d^2}=0$.
When $t \geq 18 T$, $\overline{d^2(t)}$ starts to increase and
eventually saturates to a small constant independent of $\tau$.
How fast $\overline{d^2(t)}$ relaxes to the constant is $\tau$
dependent. When $\tau=0.8$, $\overline{d^2(t)}$ almost reaches the
saturation value in one time step, suggesting rapid spreading of
the local operator for this $\tau$ value.

%%%%%%%%%%%%%%%%%%%%%%
%%%%%%%%%%%%%%%%%%

We present the dependence of $\overline{d^2(t)}$ on $x$ in
Fig.~\ref{fig:Floq_loc_d2_x}. The time for $\overline{d^2(t)}$
starting to increase is linearly proportional to the distance
between $\hat{\sigma}^x(x)$ and subsystem $A$.
%We also notice that the initial nonzero value of $\overline{d^2(t)}$ is exponentially suppressed by the distance between $\hat{\sigma}^x(x)$ and the subsystem $A$.
$\overline{d^2(t)}$ will eventually saturate to the same constant
independent of $x$. Here we provide an interpretation for this
saturation value. After long time evolution, the information of
$\hat{O}(x,-t)$ is fully scrambled in the Hilbert space and we
expect that $|\psi_1(t)\rangle$ and $|\psi_2(t)\rangle$ are close
to two independent Page states. Since the reduced density matrix
of Page state belongs to the fixed trace Wishart-Laguerre random
matrix ensemble \cite{loggasrandommatrices,
mejia_difference_2017}, we can obtain an analytical result for
$\overline{d^2(\hat{\rho}_1,\hat{\rho}_2)}$ of two independent
Page states
\begin{align}
\label{eq:analy_sat_val}
\overline{d^2(\hat{\rho}_1,\hat{\rho}_2)}=\frac{2}{m}
\end{align}
where the subsystem $A$ and $B$ have dimensions $n=2^{L_A}$,
$m=2^{L_B}$ with ratio $\alpha=n/m$ (The detail of this
calculation can be found App.~\ref{app:aver_dist}).  In
Fig.~\ref{fig:Floq_loc_com}, we present the saturation value of
$\overline{d^2(t)}$ of the Floquet model and we find that they
match well with $\overline{d^2(\hat{\rho}_1,\hat{\rho}_2)}$ of two
independent Page states. In the thermodynamic limit $L\to\infty$
while keeping $L_A$ fixed,
$\overline{d^2(\hat{\rho}_1,\hat{\rho}_2)}$ is exponentially
small. We will explain why this constant is so small in the next
section.

\begin{figure}[hbt]
\centering
 \subfigure[]{\label{fig:Floq_h} \includegraphics[width=.4\textwidth]{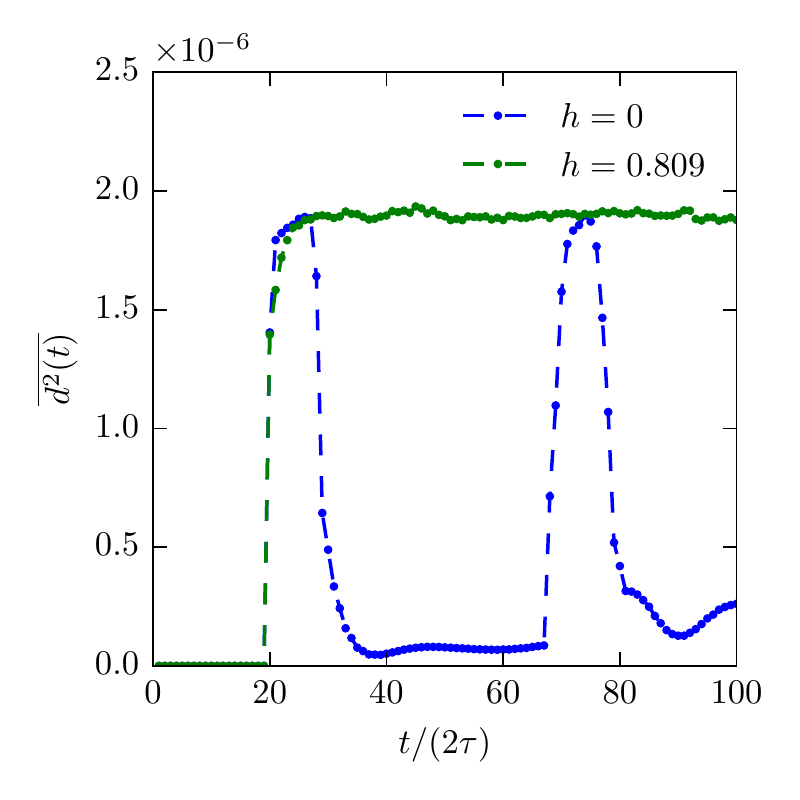}}
 \subfigure[]{\label{fig:Floq_pre} \includegraphics[width=.4\textwidth]{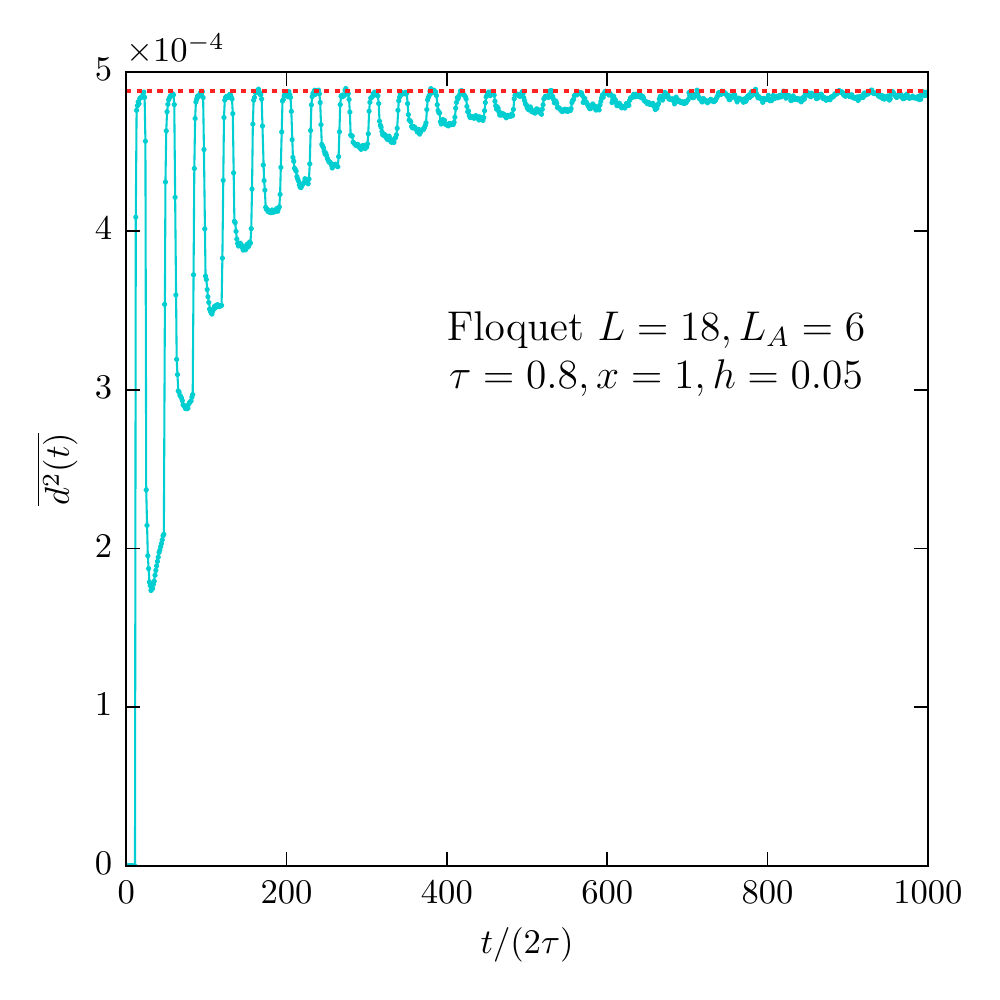}}
\caption{ (a) $\overline{d^2(t)}$ vs $t/(2\tau)$ for
non-integrable system with $h=0.809$ and integrable system with
$h=0$. (b) $\overline{d^2(t)}$ vs $t/(2\tau)$ for a system which
weakly breaks integrability with $h=0.05$. The red dashed line on
the top is $\overline{d^2}$ between two independent Page states. }
\label{fig:Floq_osc}
\end{figure}

We further study the integrable system by tuning off the field in
$z$ direction and compare it with the non-integrable system. The
early time behaviors are rather similar for both $h=0.809$
(non-integrable) and $h=0$ ($z$ field off, integrable) as shown
Fig.~\ref{fig:Floq_h}. However, in late times, $\overline{d^2(t)}$
saturates to constants in the non-integrable system while
oscillates periodically in the integrable system. The oscillation
behavior is caused by the ballistic quasi-particles created
by the local operator $\hat{O}(x)$. Once the information has
spread over the system, the quasi-particles will bounce back
and forth periodically. As a result, the length of the system and
the velocity of the quasi-particles determine the
period\cite{Cardy2007}.

We also consider $h=0.05$ which weakly breaks the integrability
and present the result in Fig.~\ref{fig:Floq_pre}. Initially, the
dynamics is similar to the case of $h=0$. As time evolves, the
amplitude of the oscillation becomes smaller and vanishes after
sufficiently long time evolution. We consider a relatively smaller
system size $L=18$ since it takes very long time for
$\overline{d^2(t)}$ to saturate. We find that the saturation value
is again the same as that for two independent Page states.
Therefore the integrability breaking term, albeit small, will
eventually wash out the memory of the initial state and drive the
system to the non-integrable behaviors.

%%%%%%%%%%%%%%%%%%%%%%
%%%%%%%%%%%%%%%%%%%%%%

%We further explicitly consider various spin chain models and find that for one dimensional spin-$\frac{1}{2}$ chain with non-local power-law interaction, $d^2(t)$ can have an appreciable {\it exponentially} increasing regime in time when the local perturbation and the subsystem $A$ are well separated.

\subsection{Quantum Ising model with a longitudinal field}

The Hamiltonian of the quantum Ising model with a longitudinal
field is
\begin{equation}
\hat H_{\rm Ising}=-\sum_i\hat\sigma^z_i\hat\sigma^z_{i+1}-h_x\sum_i \hat\sigma^x_i-h_z\sum_i \hat\sigma^z_i ~.
\label{trans_field}
\end{equation}
We take $(h_x,h_z)=(1.05,0.5)$\cite{Banuls2011} and compute
$\overline{d^2(t)}$ for Page state ensemble under unitary time
evolution governed by the Ising Hamiltonian $\hat{H}_{\rm Ising}$.
The local perturbation is still $\hat{O}(x)=\hat{\sigma^x}(x)$.

\begin{figure}[hbt]
\centering
 \subfigure[]{\label{fig:Ising_loc_d2} \includegraphics[width=.32\textwidth]{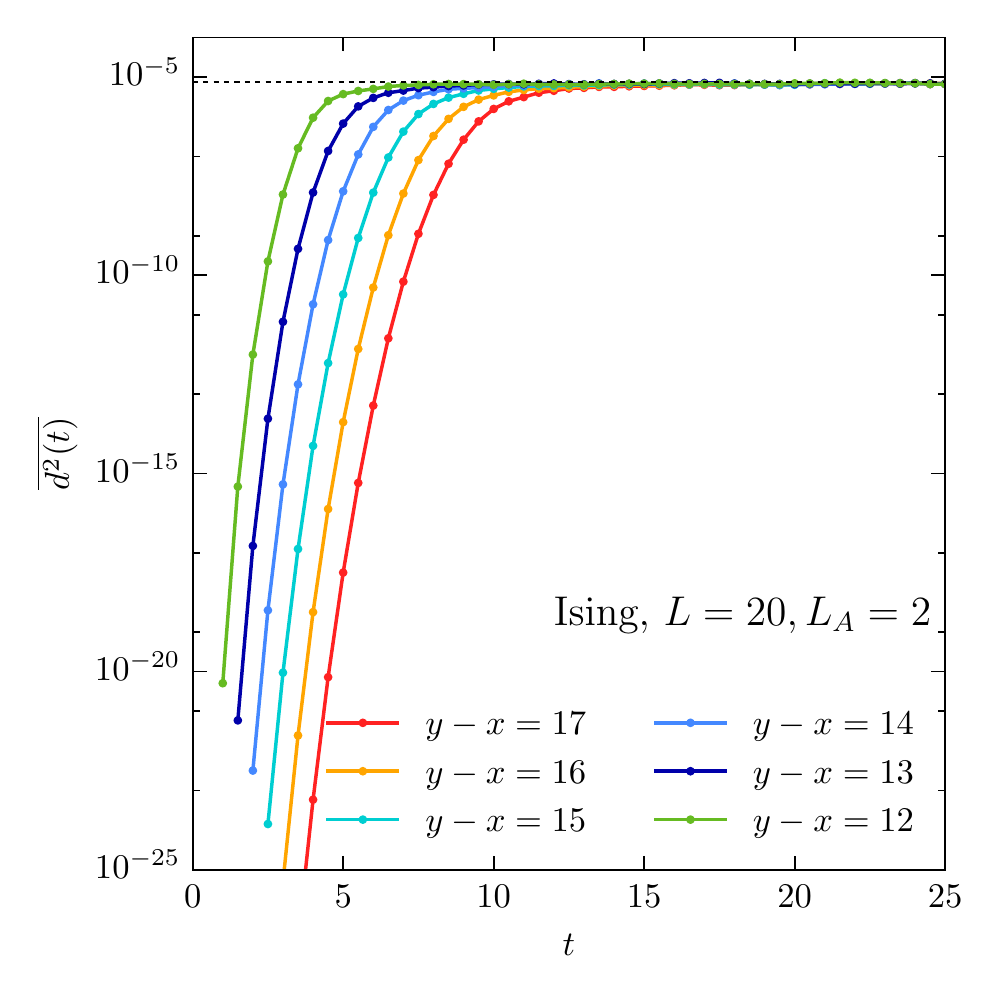}}
 \subfigure[]{\label{fig:Ising_loc_rps} \includegraphics[width=.32\textwidth]{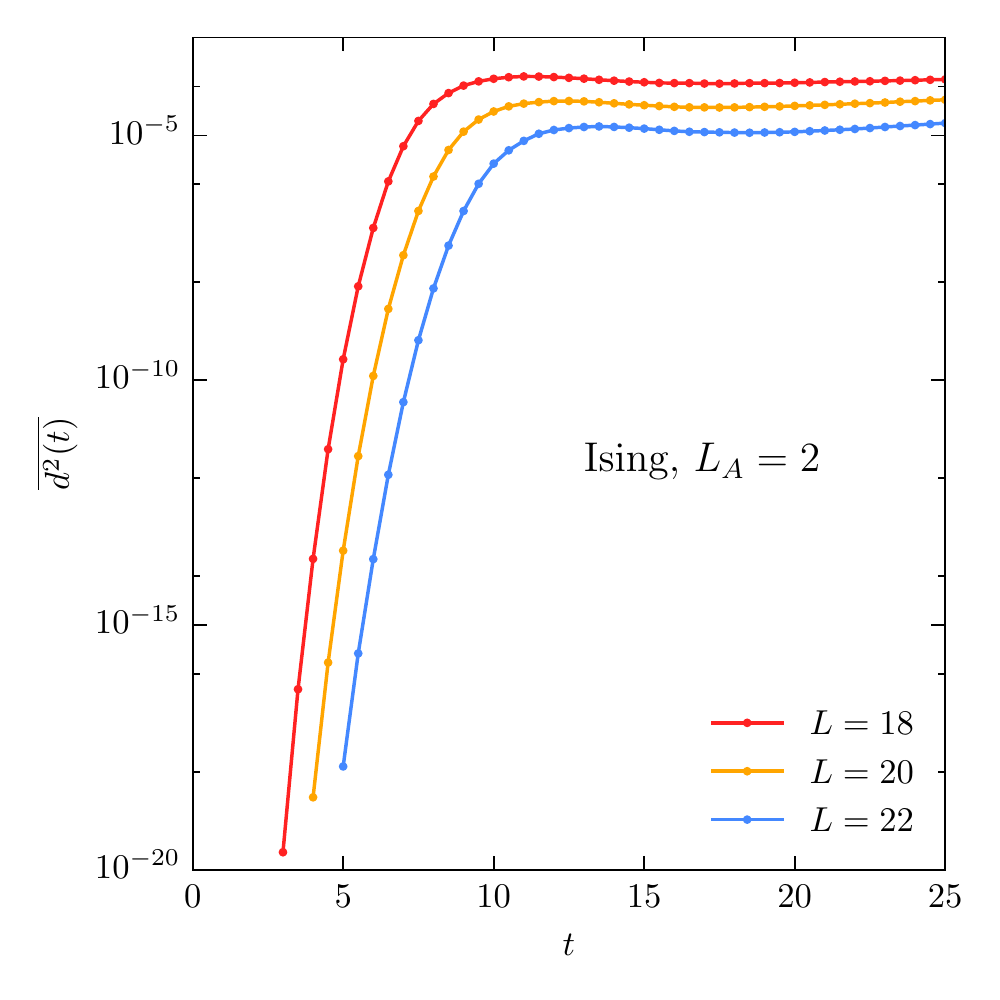}}
 \subfigure[]{\label{fig:Ising_two_rps} \includegraphics[width=.32\textwidth]{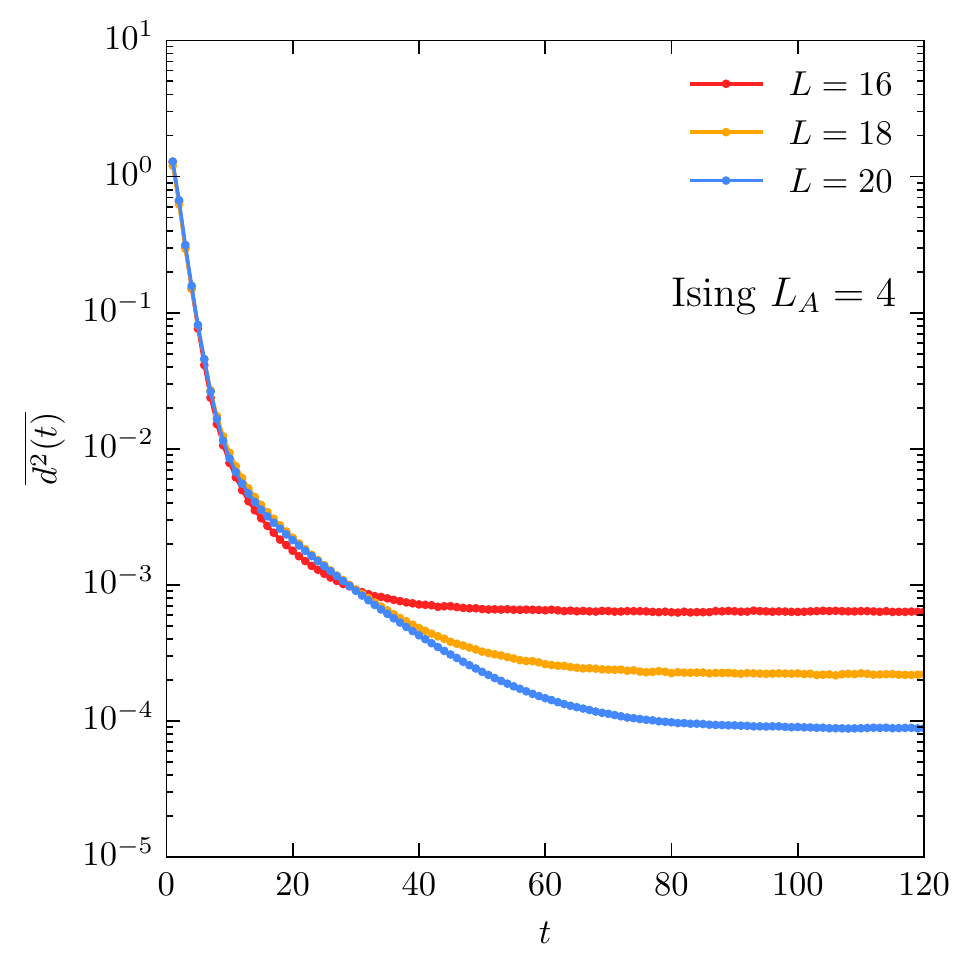}}
\caption{ (a) $\overline{d^2(t)}$ vs time $t$ for Ising model
defined in Eq.\eqref{trans_field} on the semi-log scale, where the
average is performed over 200 Page states. The dashed line is
$\overline{d^2}$ between two independent Page states. (b)
$\overline{d^2(t)}$ of the same model vs time $t$ for various $L$
on the semi-log scale, where the average is performed over 200
random product states. (c) $\overline{d^2(t)}$ vs time $t$ of two
RPS states with energy close to zero under the same unitary time
evolution.  } \label{fig:Ising_local}
\end{figure}

We present the results in Fig.~\ref{fig:Ising_local}  for various
$L_A$. As we move $\hat{O}(x)$ further away from subsystem $A$, it
takes longer time, but eventually leads to a saturation of
$\overline{d^2(t)}$ to a constant very close to values of two
independent Page states. The growth of $\overline{d^2(t)}$ is
always slower than an exponential function.  This behavior is
similar to that of the Floquet spin chain we studied in the
previous section.

%%%%%%%%%%%%%%%%%%%%%%
%%%%%%%%%%%%%%%%%%%%%%

%Here we give a physical intepretation for the small saturation of $\overline{d^2(t)}$ after sufficient time evolution. For a generic system, the initial wavefunction will eventually thermalize under its own dynamics at long times. In particular, it has been shown analytically that in $1+1$d CFT, the reduced density matrix of a small subsystem will relax to the thermal density matrix up to exponentially small corrections with the temperature determined by the initial energy \cite{Cardy2015}. For any local perturbation applied on the initial state, since the change of energy density is zero, it does not have any influence on the final thermal state. We expect it to be true in a non-integrable system which satisfies eigenstate thermalization hypothesis (ETH)\cite{Srednicki1994,Deutsch1991}.

We also explore the initial state ensemble dependence of the
saturation value. For the case of Ising model, we prepare another
set of states that are random product states (RPS) with energy
$E\in[-1,1]$. A RPS is a tensor product of state on each site
whose spin directions are uniformly distributed on Bloch sphere.
It thus has zero initial entanglement entropy (EE). We apply a
perturbation on the RPS state and then compute the time dependent
distance $\overline{d^2(t)}$. The results are presented in
Fig.~\ref{fig:Ising_loc_rps}. The saturation values of
$\overline{d^2(t)}$ for this set of initial states is different
from the ones of two random states, but it still decays
exponentially as we increase the system size.

The small saturation value of $\overline{d^2(t\to\infty)}$ can be
understood in the following way. For a generic system, the initial
wavefunction will eventually thermalize under its own dynamics at
long times \cite{Srednicki1994,Deutsch1991}. In particular, it has
been proven that in $1+1$d conformal field theory (satisfying
generalized Gibbs ensemble \cite{Rigol2006, Rigol2007}), the
reduced density matrix of a small subsystem will relax to the
thermal density matrix up to exponentially small corrections with
the temperature determined by the initial energy density
\cite{Cardy2015}. Therefore, if the two initial states have the
same energy density, after sufficiently long time evolution,
the reduced density matrix of both states will thermalize at the
same effective temperature with $\overline{d^2(t)}$ approaching to
a small constant.
%\rev{The Ising model considered here satisfies ETH\cite{Banuls2011}. With the local perturbation applied to $|\psi_2 \rangle$, its energy density will approach and eventually becomes the same as the energy density of $|\psi_1\rangle$ if $L \rightarrow \infty$.}
The Ising model considered here satisfies ETH\cite{Banuls2011}.
The local perturbation applied to $|\psi_2 \rangle$ will not
change energy density and gives rise to an exponentially small
saturation value of $\overline{d^2(t\to\infty)}$.

Furthermore, we study the saturation value of $d^2(t)$ when
$|\psi_1\rangle$ and $|\psi_2\rangle$ are not connected by a local
unitary transformation but have the same energy density. We choose
$|\psi_1\rangle$ and $|\psi_2\rangle$ to be two different PRS
states. At $t=0$, $\overline{d^2(t)}$ for subsystem $A$ is a $O(1)$ constant. After
sufficiently long time evolution, $\overline{d^2(t)}$ relaxes to a
small constant, which decreases exponentially as $L$ increases and
is also consistent with ETH. 
%{\rd Questions: in this paragraph, we
%are not using reduced density matrix, but the density matrix of
%he whole system, right? If so, we should probably point this
%out.}

\section{The spin chain models with power-law interaction}
\label{power_law_interaction}
\subsection{Floquet spin-$\frac{1}{2}$ chain model}

In this section, we consider Floquet model with power-law
interaction
\begin{align}
\hat{U}_F=\exp[-i\tau \hat{H}_z]\exp\left[-i\tau\hat H_x\right] ~,
\label{flo_op_power}
\end{align}
where
\begin{align}
&\hat{H}_x=\sum_{j=1}^{L} g\hat\sigma_j^x\nonumber\\
&\hat{H}_z=\sum_{j>k}\frac{\hat\sigma_j^z\hat\sigma_{k}^z}{(j-k)^\alpha}+\sum_{j=1}^Lh\hat\sigma_j^z ~.
\end{align}
We fix parameter  $(g,h,\tau)=(0.9045, 0.8090,0.4)$ and compute
$\overline{d^2(t)}$ averaged over the Page state ensemble. We take the power law exponent $\alpha=2.0$.

\begin{figure}[hbt]
\centering
 \subfigure[]{\label{fig:Floq_pow_d2_x} \includegraphics[width=.4\textwidth]{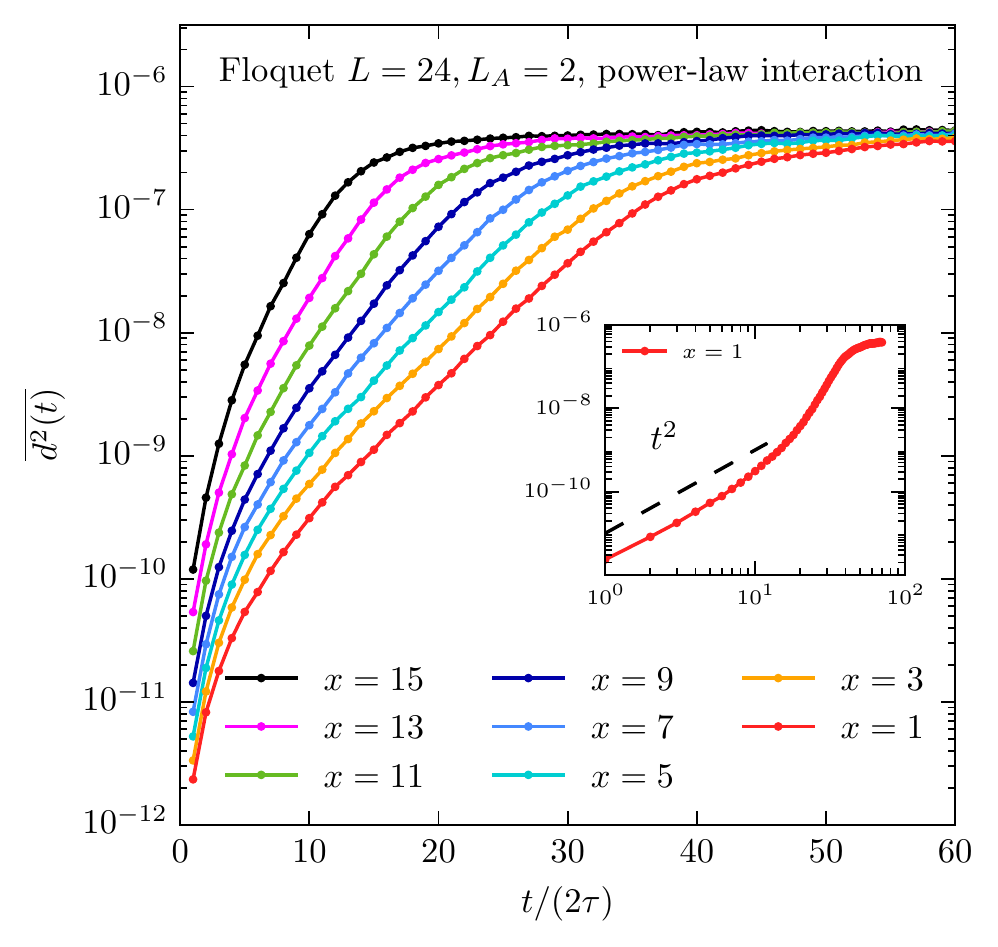}}
 \subfigure[]{\label{fig:Floq_pow_d2_LA} \includegraphics[width=.4\textwidth]{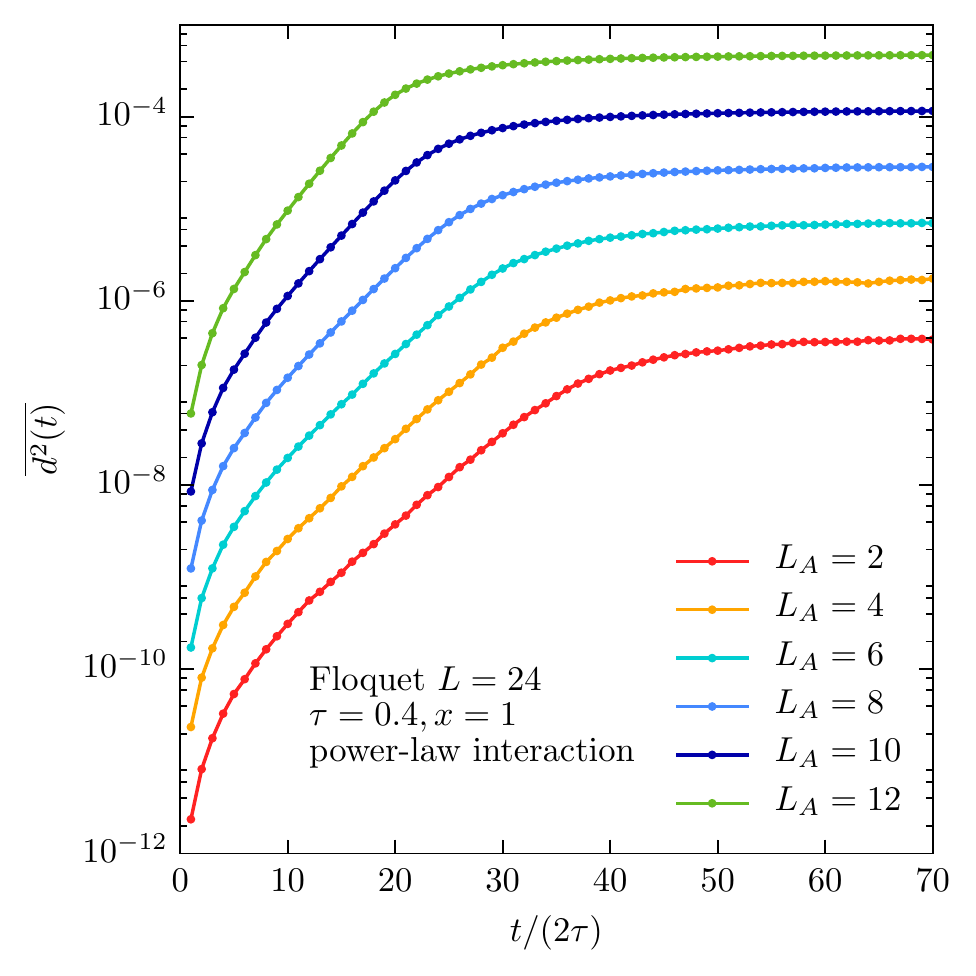}}
\caption{ (a) $\overline{d^2(t)}$ vs $t/(2\tau)$ for various $x$
with fixed $L_A$ on the semi-log scale for Floquet operator in
Eq.\eqref{flo_op_power}. The average is over $200$ Page states.
(b) $\overline{d^2(t)}$ vs $t/(2\tau)$ for various $L_A$ with
fixed $x$ on the semi-log scale. We perform $200$ Pages states
average for $L_A=2$ and $8$ Page states average for $L_A \geq 4$.
% For $L_A=2$, the average is over $200$ Page states, while for $L_A\geq 4$, the average is performed over $8$ Page states.
}
\label{fig:Floq_pow_d2}
\end{figure}

As shown in Fig.~\ref{fig:Floq_pow_d2_x}, when the distance
between region $A$ and $\hat{\sigma}^x(x)$ is large, we can
observe a clear intermediate fast growing regime. It has positive
curvature on the log-log scale in the inset, indicating that it is
faster than any power law growth. The fact that it is a straight
line on the semi-log plot confirms the existence of an exponential
growth regime.

This is different from the spin chain models with local
interaction, where an exponentially growing regime is
invisible in the parameter range we choose. The physical
interpretation for this regime will be given in
Sec.~\ref{op_scramble}.

Notice that at early time, $\overline{d^2(t)}$ scales as $t^2$
(inset of Fig.~\ref{fig:Floq_pow_d2_x}), which can be understood
by taking short time expansion for $\hat{\sigma}^x(x,-t)$. Similar
behavior is also found in Ref.~\onlinecite{Dora2017} and we will
not discuss it here.

For fixed $L_A$, as we decrease the distance between
$\hat{\sigma}^x(x)$ and subsystem $A$, this intermediate
exponentially growing regime becomes smaller and  vanishes
eventually. Nevertheless, $\overline{d^2(t)}$ always saturates to
the same constant, which is also the same as that of two
independent Page states. We also present $\overline{d^2(t)}$ for
various $L_A$ with fixed $x=1$ in Fig.~\ref{fig:Floq_pow_d2_LA}.
As we increase the subsystem length $L_A$, the distance between
$x$ and the subsystem $A$ reduces, and therefore the exponentially
growing regime becomes smaller. We observe that the growth rate is
not very sensitive to the distance between $\hat{\sigma}^x(x)$ and
region $A$.

Notice that when $L_A=2$, $\overline{d^2(t)}$ is averaged of 200
Page states due to large fluctuation in the data. When the
subsystem has $L_A\geq 4$, the fluctuation is much smaller and the
ensemble average is unnecessary. In the numerical calculation, we
take average over 8 states.

%%%%%%%%%%%%%%%%%%%%%%
%%%%%%%%%%%%%%%%%%

\subsection{Ising model with power-law interaction}

We now consider a static Hamiltonian with a non-local power-law
interaction,
\begin{equation}
\hat H=-\sum_{j>k}\frac{\hat\sigma^z_j\hat\sigma^z_{k}}{(j-k)^{\alpha}}-h_x\sum_i \hat\sigma^x_i-h_z\sum_i \hat\sigma^z_i ~.
\label{Ising_power}
\end{equation}
where we fix the parameter $(h_x,h_z)=(1.05, 0.5)$ to be the same
as in Eq.\eqref{trans_field}. We take the power law exponent
$\alpha=2.1$, where we find the longest exponentially growing
regime for $\overline{d^2(t)}$.

\begin{figure}[hbt]
\centering
 \subfigure[]{\label{fig:Ising_pow_d2_x} \includegraphics[width=.4\textwidth]{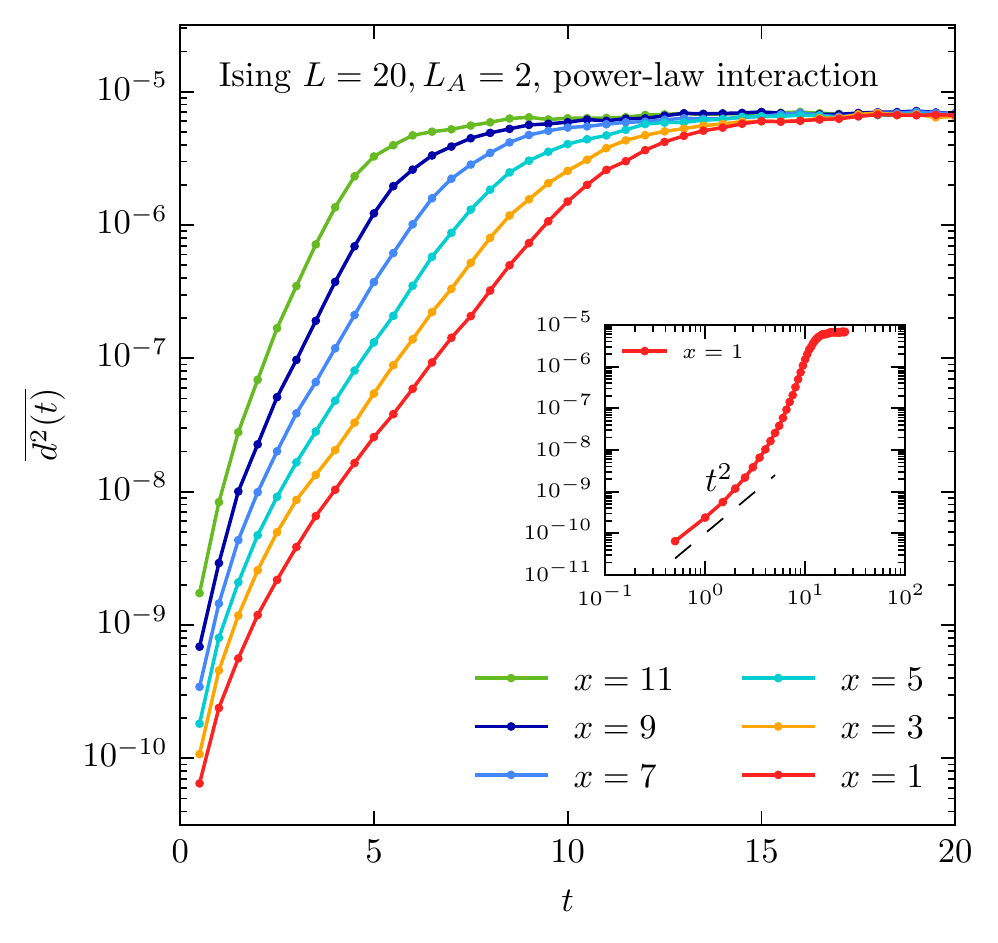}}
 \subfigure[]{\label{fig:Ising_pow_d2_LA} \includegraphics[width=.4\textwidth]{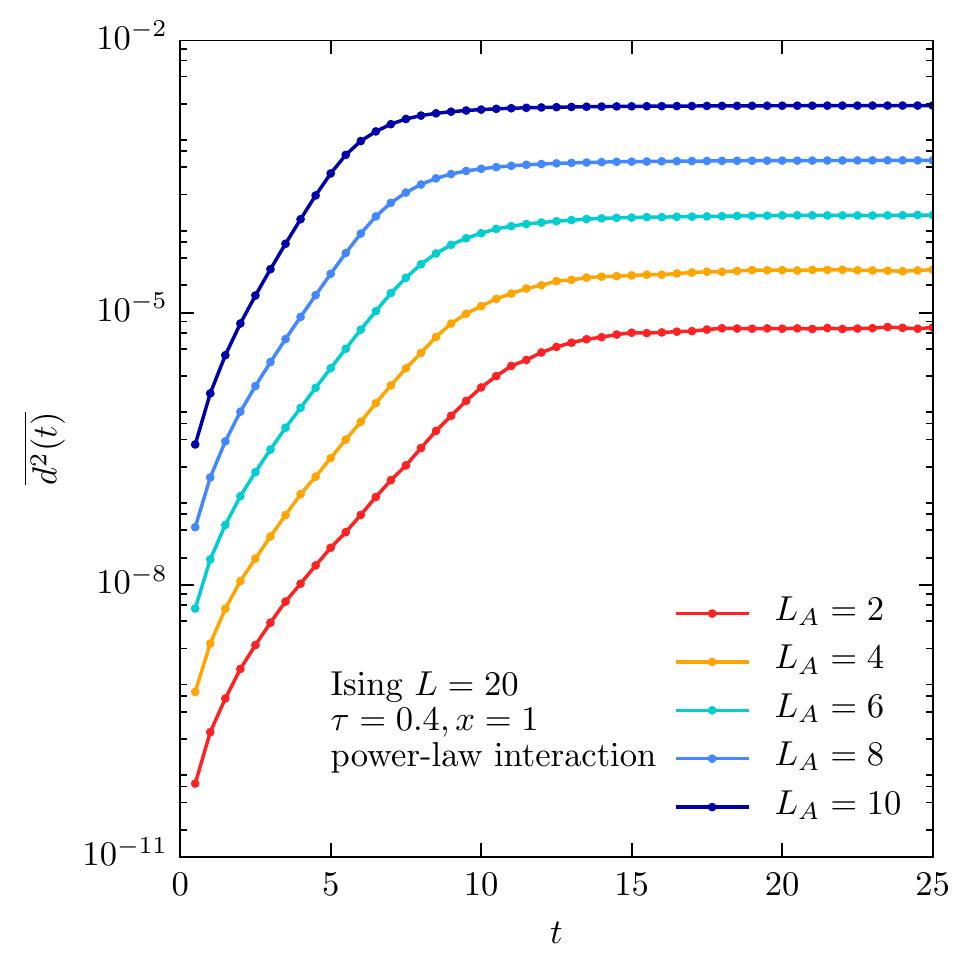}}
\caption{ (a) $\overline{d^2(t)}$ vs $t$ for various $x$ with
fixed $L_A$ on the semi-log scale for Ising model  in
Eq.\eqref{Ising_power}. The average is over $200$ Page states. (b)
$\overline{d^2(t)}$ vs $t$ for various $L_A$ with fixed $x$ on the
semi-log scale. We perform $200$ Pages states average for $L_A=2$
and $8$ Page states average for $L_A \geq 4$.
% For $L_A=2$, the average is over $200$ Page states, while for $L_A\geq 4$, the average is performed over $8$ Page states.
}
\label{fig:Ising_pow_d2}
\end{figure}

We still take $|\psi_1\rangle$ as a Page state and
$|\psi_2\rangle=\hat{\sigma}^x(x)|\psi_1\rangle$. As we can
observe in Fig.~\ref{fig:Ising_pow_d2}, the non-local power-law
interaction leads to an exponentially growing regime in
$\overline{d^2(t)}$, which becomes longer as we increase the
distance between $\hat{\sigma}^x(x)$ and region $A$. The
saturation value of $\overline{d^2(t)}$ is still the same as that
for two independent Page states.

%%%%%%%%%%%%%%%%%%%%%%
%%%%%%%%%%%%%%%%%%

\section{Connection between $d^2(t)$ and the square of commutator}
\label{op_scramble}

In this section, we use the operator spreading picture to
interpret the behaviors of $d^2(t)$ and relate it to the square of
commutator. We will show that they are essentially the same
quantity after a proper rescaling in the models we consider.

%both of them can detect the surge of chaos through the quantum Lyapunov exponent.

\subsection{$d^2(t)$ in operator spreading picture}

For simplicity, we will restrict our discussion to the
spin-$\frac{1}{2}$ chain that is numerically studied in this
paper, but a generalization to models with $q>2$ dimensional
on-site Hilbert space is straightforward.

On each site of the spin-$\frac{1}{2}$ chain, we can define a
local operator basis which consists of the identity operator
$\hat{I}_2$ and the three Pauli matrices $\hat{\sigma}^x$,
$\hat{\sigma}^y$, $\hat{\sigma}^z$. Using the tensor product of
the local basis, we can build a complete operator basis $\{
\hat{H}_j\}$ for the entire Hilbert space, in which each operator
$\hat{H}_j$ is a string of $\hat{\sigma}^{x,y,z}$ and $\hat{I}_2$.
This basis has a natural inner product
$\mbox{Tr}(\hat{H}_i\hat{H}_j)=\delta_{ij}|H|$, where $|H|$ is the
size of the Hilbert space. We can choose $4^{L_B}$ operators
$\{B_j\}$ from $\{H_j\}$ which consists of only identity operators
in region $A$. They form an operator basis completely confined in
region $B$. The average over these operators can be used to take a
partial trace on any operator $\hat{O}$ in the full Hilbert
space~\cite{Hosur_2016},
\begin{align}
\frac{1}{|B|}\sum_i \hat{B_i}\hat{O}\hat{B}_i= \mbox{Tr}_B(\hat{O}) \otimes \hat{I}_B
\end{align}
where $|B|$ is the Hilbert space dimension of region $B$. A
corollary of this identity is that for any operator $\hat{O}$
\begin{equation}
\label{eq:OTO-EE}
\Tr_A ( \Tr_B \hat{O}  \Tr_B \hat{O} ) = \Tr_A \Tr_B ( ( \Tr_B \hat{O} \otimes I_B )  \hat{O} ) = \frac{1}{|B|}\sum_{i} \Tr( B_i \hat{O} B_i   \hat{O})
\end{equation}
This equation has been used to show the connection between the OTO
correlator and operator entanglement entropy \cite{Hosur_2016,
Fan_2016, harrow_random_2009}.
%
%%%%%%%%%%%%%%
\begin{figure}%[hbt]
\centering
\includegraphics[width=.6\textwidth]{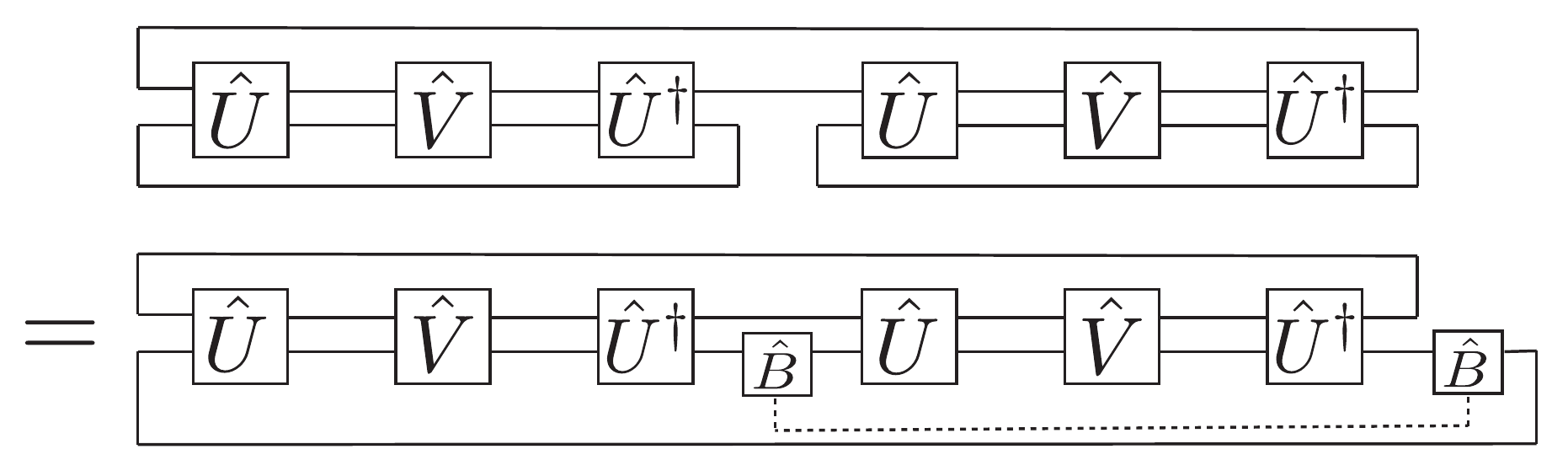}
\caption{Diagrammatic proof of Eq.\eqref{eq:OTO-EE}. The top
channel corresponds to the region $A$ and the bottom channel
corresponds to the region $B$. Dotted line indicates summation
over $\hat{B}$ operators.  } \label{fig:d2_diagram}
\end{figure}
%%%%%%%%%%%%%%

Using the above identity, we are going to rewrite $d^2(t)$ in
terms of spreading of local operator in real space. For the pure
state under unitary time evolution, the reduced density matrix for
region A (after tracing out region B) takes the form,
\begin{align}
\hat{\rho}(t)=\mbox{Tr}_B\left[\hat{U}(t)|\psi\rangle\langle
\psi|\hat{U}^{\dag}(t)\right]
\end{align}
The difference of the time dependent density matrices in region
$A$ is $\hat{\rho}_1(t ) - \hat{\rho}_2(t) =  \hat{U}(t) \hat{V}
\hat{U}^{\dagger} (t)$, where
$\hat{V}=|\psi_1\rangle\langle\psi_1|-|\psi_2\rangle\langle\psi_2|$.
Therefore the time dependent distance is
\begin{align}
d^2(t)=\mbox{Tr}_A\left[\mbox{Tr}_B\hat{U}(t)\hat{V}\hat{U}^{\dag}(t)\right]^2=\frac{1}{|B|}\sum_{i}\mbox{Tr}\left[\hat{B}_i(t)\hat{V}\hat{B}_i(t)\hat{V}\right]
\label{d_2}
\end{align}
where $\hat{B}_i(t) = \hat{U}^{\dag}(t)\hat{B}_i \hat{U} $. The
above equation is obtained by using the operator identity in
Eq.\eqref{eq:OTO-EE} and is also diagrammatically interpreted in
Fig.~\ref{fig:d2_diagram}.

By plugging the explicit form of $\hat{V}$ into Eq.\eqref{d_2}, we
can expand $d^2(t)$ as
\begin{align}
d^2(t)=\frac{1}{|B|}\sum_i\left[ \langle \psi_1|\hat{B}_i(t)|\psi_1\rangle^2+\langle \psi_2|\hat{B}_i(t)|\psi_2\rangle^2
-2\langle\psi_1 | \hat{B}_i(t)|\psi_2\rangle\langle\psi_2 |\hat{B}_i(t)|\psi_1\rangle  \right]
\end{align}

Take $|\psi_1\rangle$ as a Page state, we can use its properties
(Eq.\eqref{identity_o} and Eq.\eqref{identity_oo} in
App.~\ref{app:aver_over_page}) and obtain the following
expression,
\begin{equation}
\label{eq:d2_op_scramble}
\begin{aligned}
\overline{d^2(t)}&=\frac{2}{|B|}\sum_i \frac{ \Tr^2 ( \hat{B}_i ) - \Tr^2 ( \hat{B}_i \hat{O}( x, -t ) )}{|H|(|H| + 1)} \\
&= \frac{2}{|B|}\frac{|H|}{|H|+1}\left\{ 1- \frac{1}{|H|^2}\sum_i\mbox{Tr}^2\left[ \hat{B}_i\hat{O}(x,-t) \right] \right\}
\end{aligned}
\end{equation}
This is one of the most important results in this paper. Notice
that this averaged result is state independent and depends only on
the spreading of $\hat{O}(x, -t)$ in the Hilbert space.
Specifically, $\hat{O}(x,-t)$ can always be expanded in the
operator basis as
\begin{equation}
\hat{O}(x,-t)=\sum_j \alpha_j(t) \hat{H}_j
\label{O_coeff}
\end{equation}
For Hermitian operators on Hermitian basis $\{\hat{H}_j\}$, the
components $\alpha_j(t)$ are real numbers and they satisfy the
normalization constraint $\sum_j\alpha_j(t)^2=1$. The trace with
the basis $\{\hat{B}_i\}$ evaluates the component of
$\hat{O}(x,-t)$ that is completely confined in region $B$.
Therefore $\overline{d^2(t)}$ is proportional to the weights of
the operator basis that is not entirely in region $B$.

The spreading of operator  $\hat{O}(x,-t)$ depends on the behavior
of the coefficients $\alpha_j(t)$ and gives rise to an effective
hydrodynamical description in one dimensional chaotic systems
\cite{Nahum2017, Keyserlingk2017, Khemani2017,Rakovszky2017}. As
time evolves, the string operators dominating the sum in
Eq.\eqref{O_coeff} grow spatially and we can define the end point
distribution to measure the spreading of $\hat{O}(x,-t)$,
\begin{equation}
f_R(s,t)\equiv \sum_j \alpha_j(t)^2 \delta( \text{right most site}(\hat{H}_j ) = s  )
\end{equation}
which counts all the components of the basis that ends at site $s$
at time $t$. It is subjected to the normalization
\begin{equation}
\int_{0}^{L} f_R( s, t ) ds = 1
\end{equation}
and can be interpreted as the (coarse grained) probability
distribution of operators in $\hat{O}(-x, t)$ that ends at
location $s$ at time $t$. One can similarly define the left end
distribution $f_L(s, t)$.

\begin{figure}%[hbt]
\centering
\includegraphics[width=.5\textwidth]{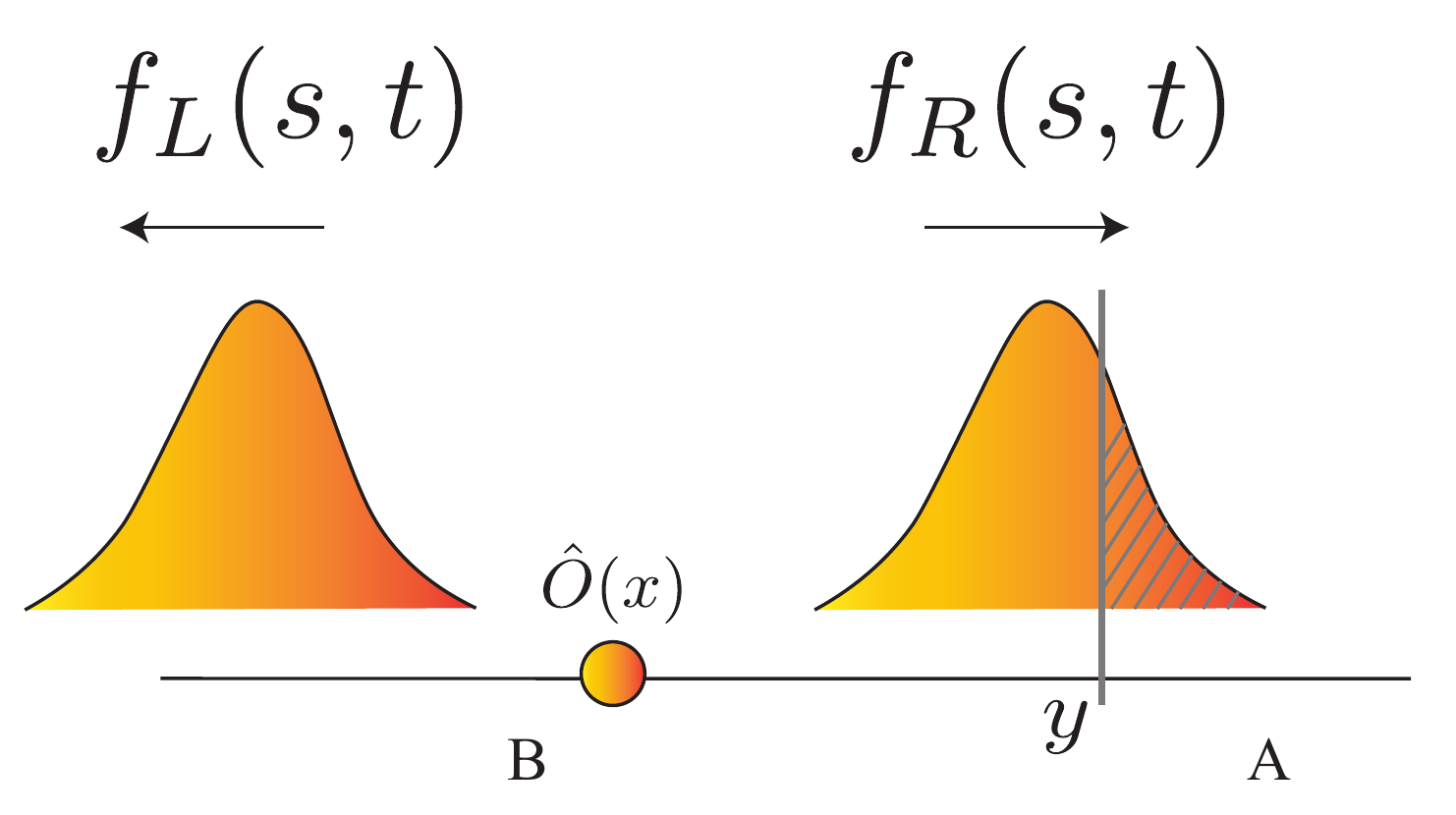}
\caption{The schematics of the wave front $f_L(s,t)$ and
$f_R(s,t)$ of $\hat{O}(x,-t)$. } \label{fig:op_scramble}
\end{figure}

Initially, the local operator $\hat{O}(x)$ only occupies one
operator basis on one site. In early time, almost all operators
contained in $\hat{O}(x, -t)$ are confined in region $B$, hence
$\overline{d^2(t)}$ is small. Under the chaotic evolution, the end
points distribution should move outwards on the left and right
ends. Fig.~\ref{fig:op_scramble} is the schematic of the shape of
the distribution at late times. If the right end points of most
operators are moving to the right, the distribution $f_R(s, t)$
should have a sharp wave front. Once the wave front crosses the
cut at $y$, there will be an appreciable fraction of
operators in $\hat{O}(x,-t)$ that has spread into region $A$,
which contributes to the rapid growth of $d^2(t)$. In fact, the
area of the wave front cut by the boundary of $A$ and $B$ is
proportional to $\overline{d^2(t)}$,
\begin{equation}
\overline{d^2(t)} = \frac{2}{|B|} \frac{|H|}{( |H| + 1 )}
\int_{y}^{L} f_R( s, t ) ds  \sim \frac{2}{|B|} \int_{y}^{L} f_R(
s, t ) ds
\end{equation}
After sufficiently long time evolution, we expect that all the
string operators in $\hat{O}(x, -t)$ have spanned the entire
system. This means that $f_R(s,t)$ is localized at $s = L$ and
$\overline{d^2(t)}$ saturates to
\begin{equation}
\overline{d^2(t)} \sim \frac{2}{|B|}
\end{equation}
This is consistent with analytical result of the distance of two
independent Page states (see App.~\ref{app:aver_dist})

%%%%%%%%%%%%%%
%%%%%%%%%%%%%%

\subsection{$d^2(t)$ and square of commutator}

\begin{figure}[hbt]
\centering
 \subfigure[]{\label{fig:Floq_local_d2_oto} \includegraphics[width=.38\textwidth]{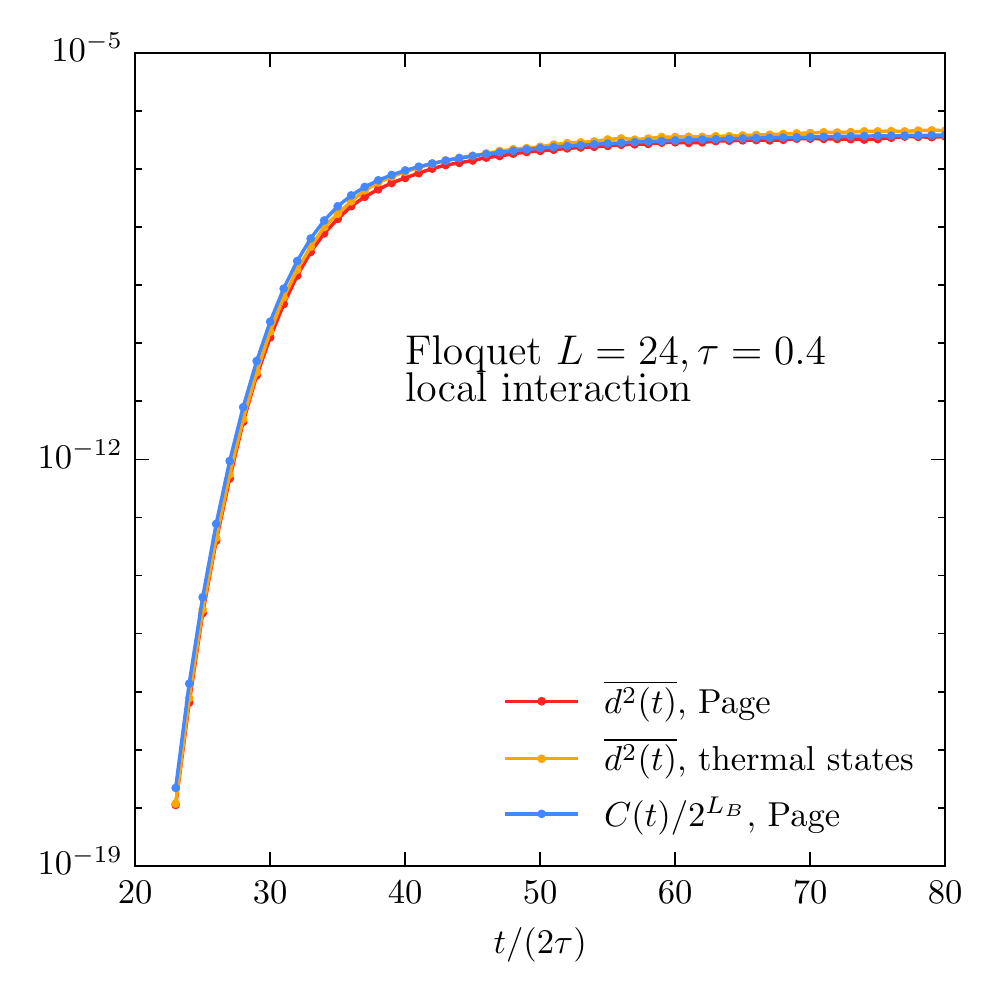}}
 \subfigure[]{\label{fig:Floq_pow_d2_oto} \includegraphics[width=.38\textwidth]{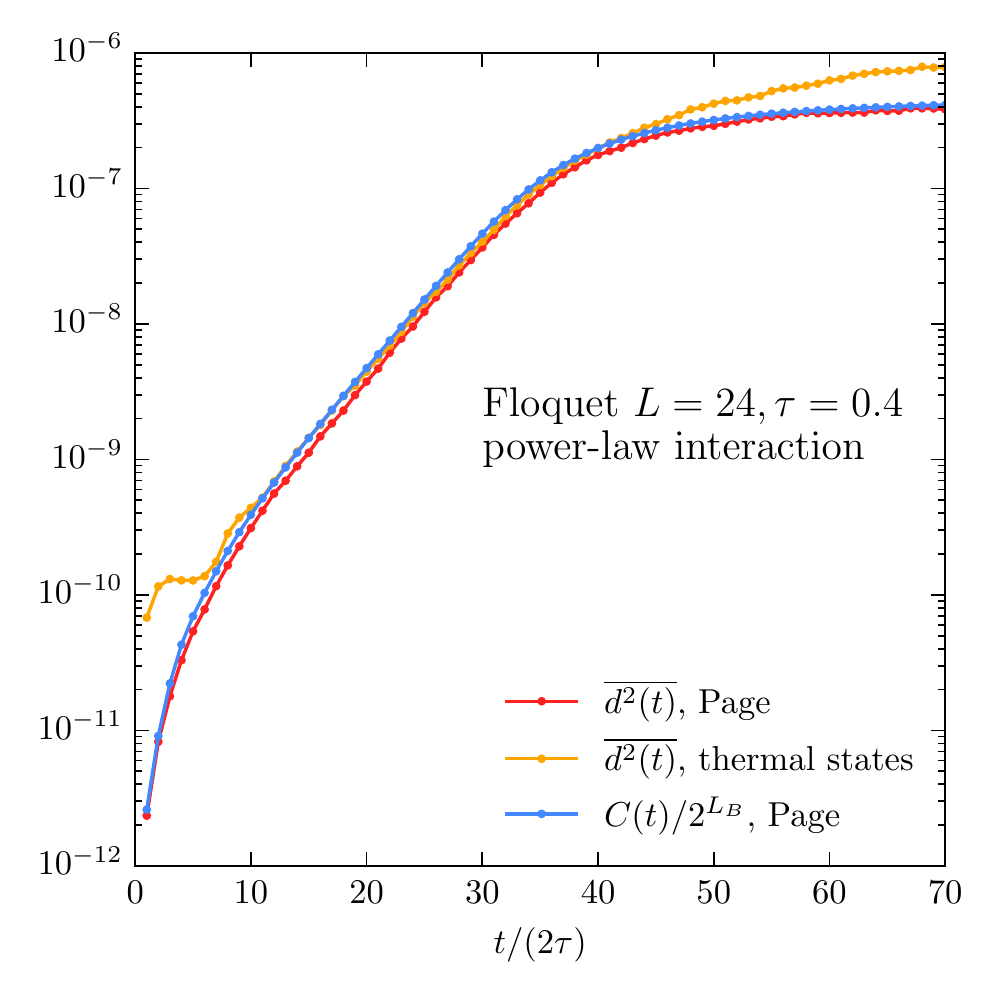}}
  \subfigure[]{\label{fig:Ising_local_d2_oto} \includegraphics[width=.38\textwidth]{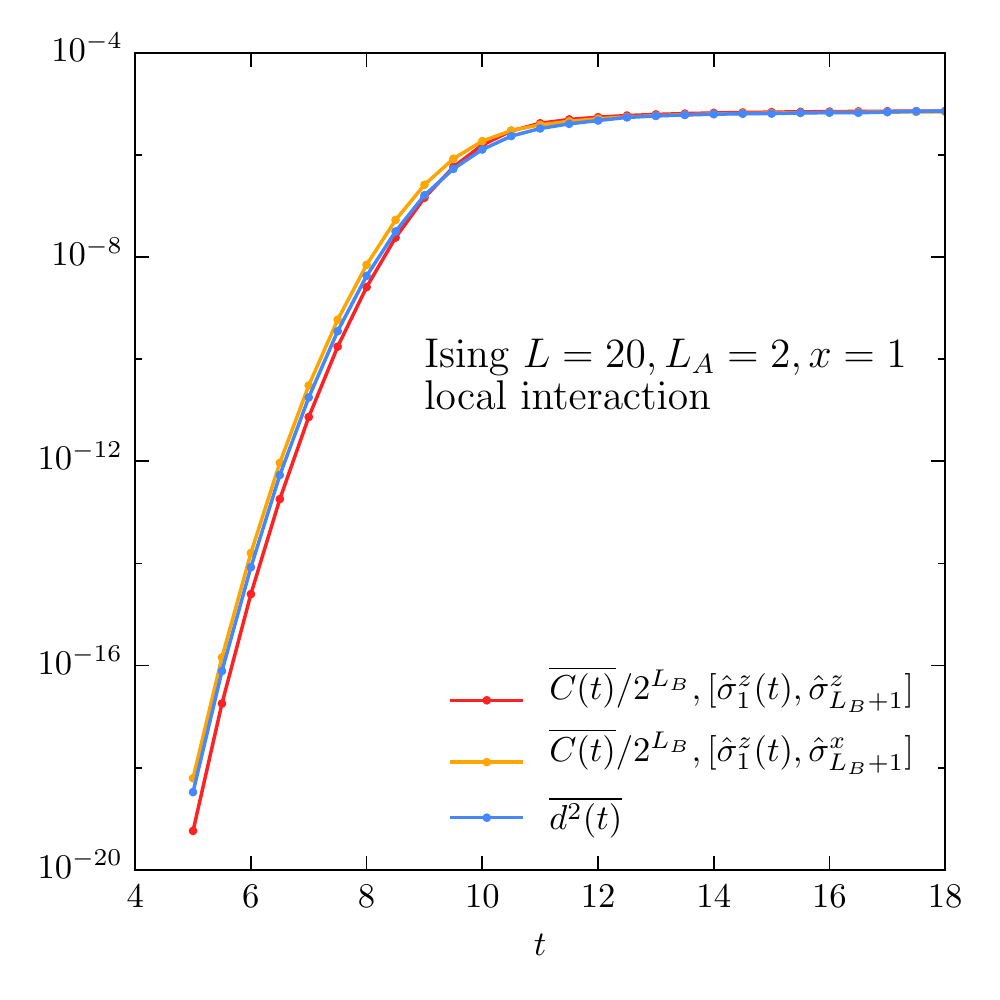}}
 \subfigure[]{\label{fig:Ising_pow_d2_oto} \includegraphics[width=.38\textwidth]{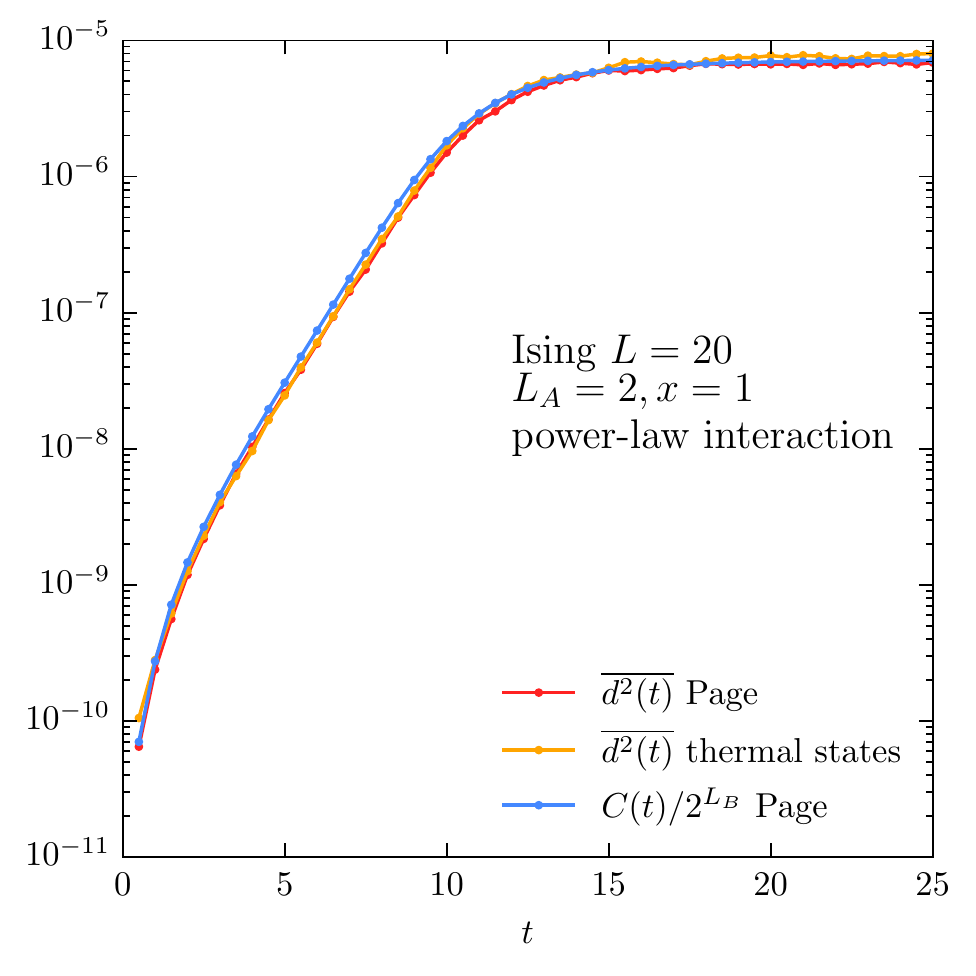}}
\caption{ (a) $\overline{d^2(t)}$ and $\overline{C_s(t)}$ vs
$t/(2\tau)$  on the semi-log scale for Floquet operator in
Eq.\eqref{flo_op} with local interaction ($L_A=2$). (b) $\overline{d^2(t)}$ and $\overline{C_s(t)}$
vs $t/(2\tau)$  on the semi-log scale for Floquet operator in
Eq.\eqref{flo_op_power} with power-local interaction ($L_A=2$). The slope of the curve (for the exponentially growing regime) is $0.244$. (c)
$\overline{d^2(t)}$ and $\overline{C_s(t)}$ vs $t$  on the
semi-log scale for the static Hamiltonian in
Eq.\eqref{trans_field} with local interaction. (d)
$\overline{d^2(t)}$ and $\overline{C_s(t)}$ vs $t$  on the
semi-log scale for the static Hamiltonian in
Eq.\eqref{Ising_power}.  The slope of the curve (for the exponentially growing regime) is $0.902$. In (a),(b) and (d), we take $\hat{O}(x,t=0)=\hat\sigma^x$ in
$\overline{d^2(t)}$ and $\hat{O}(x,t=0)=\hat{O}^\prime(y)=\hat\sigma^x$ in
$\overline{C_s(t)}$. In (c), for $\overline{C_s(t)}$, we fix $\hat{O}(x,t=0)=\hat\sigma^z$ and consider $\hat{O}^\prime(y)=\hat\sigma^z$ (red curve) and $\hat{O}^\prime(y)=\hat\sigma^x$ (yellow curve). The difference between these two curves at early time is because $\sigma^z$ has a larger overlap with energy density operator. The $\overline{d^2(t)}$ of  $\hat{O}(x,t=0)=\hat\sigma^z$ (dashed black curve) lies in the middle between the red and blue curves.}
\label{fig:pow_d2_oto}
\end{figure}

\begin{figure}[hbt]
\centering
 \subfigure[]{\label{fig:Floq_d2_OTO_L_A} \includegraphics[width=.4\textwidth]{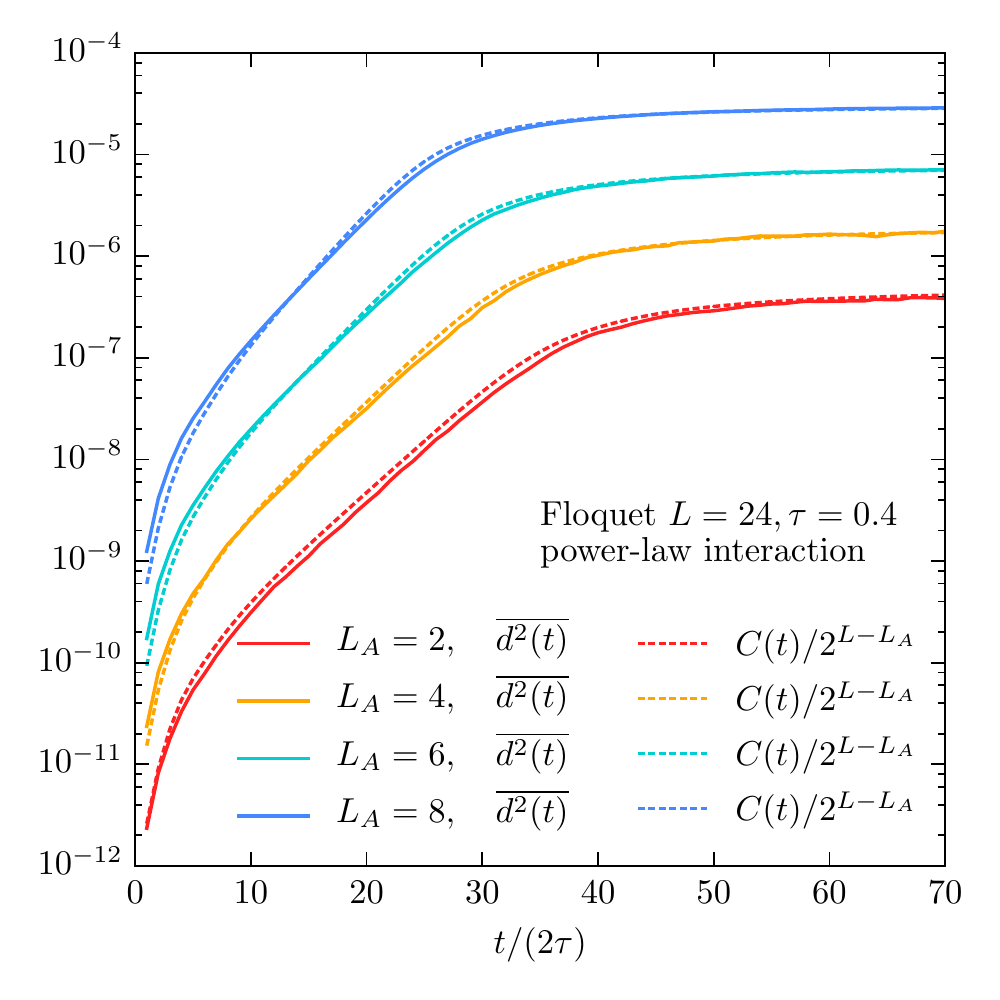}}
 \subfigure[]{\label{fig:Floq_wavefront} \includegraphics[width=.4\textwidth]{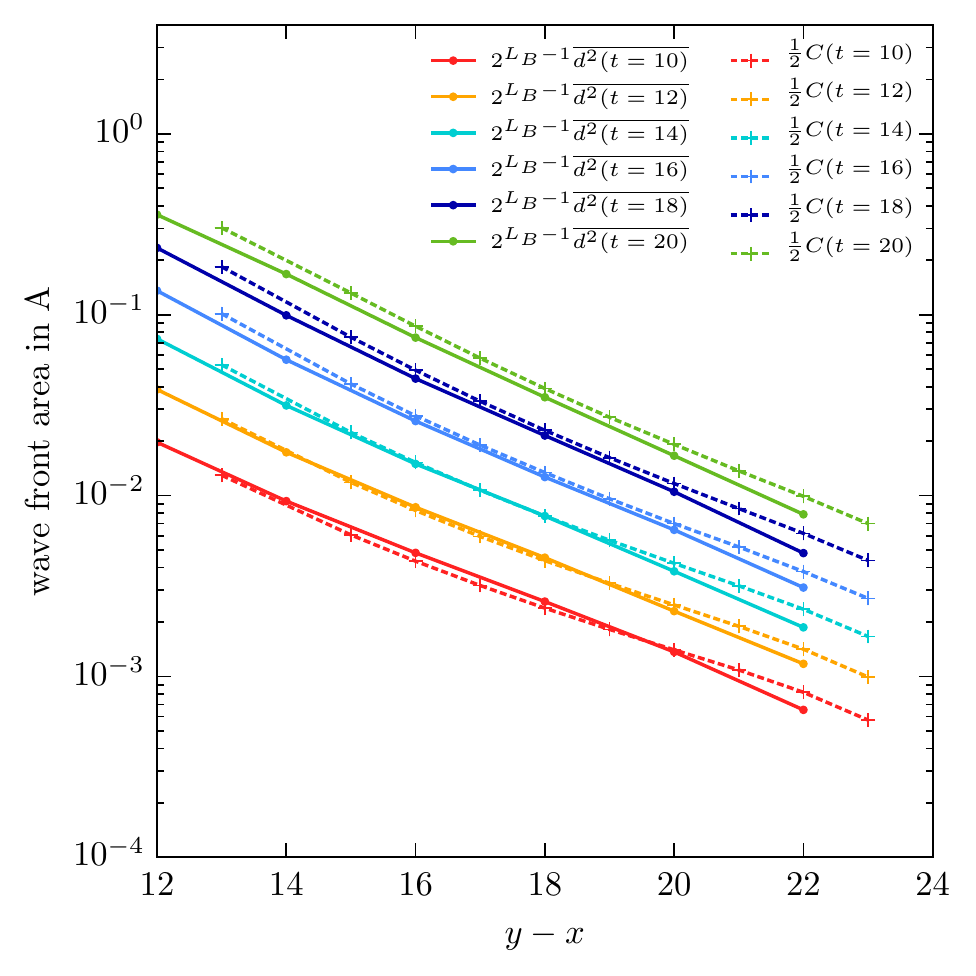}}
\caption{ (a) $\overline{d^2(t)}$ and $\overline{C_s(t)}$ vs
$t/(2\tau)$  on the semi-log scale for Floquet operator in
Eq.\eqref{flo_op_power} for various $y$ and fixed $x=1$ on the
semi-log scale. (b) $2^{L_B-1}\overline{d^2(t)}$ (rescaled $d^2$)
and $\overline{C_s(t)}/2$ vs the distance $|y-x|$  for various
time $t$. Both quantities are measuring the area of the wave front
in region $A$. } \label{fig:pow_tail}
\end{figure}

We would like to compare $d^2(t)$ with another interesting
quantity, the square of commutator
\begin{equation}
C(t) = -\frac{1}{|H|}\Tr [\hat{O}(x,-t),\hat{O}^{\prime}(y)]^2
\label{C_trace}
\end{equation}
where we take $\hat{O}^{\prime}(y)$ as a local Pauli operator
sitting at position $y$ that does not evolve. We will just take it
be $\hat\sigma^x$. In the operator basis we choose, only those do
not commute with $\hat{O}^\prime(y)$ will contribute to $C(t)$,
i.e.,
\begin{equation}
C(t) = - \sum_j  \alpha_j^2(t) \frac{1}{|H|} \Tr [\hat{H}_j, \hat{\sigma}^x(y ) ]^2
= 4  \sum_j  \alpha_j^2(t)  \delta( \hat{H}_j\big|_y = \hat{\sigma}^{y,z} )
\end{equation}
Under the chaotic evolution, it is reasonable to assume that the
probability of having $\hat{\sigma}^{x,y,z}$ and $\hat{I}$ are
essentially the same \footnote{This assumption may not be valid in the presence of conservation law (See Fig.~\ref{fig:Ising_local_d2_oto}). Moreover, it may also break down in the chaotic system which weakly breaks integrability. All these problems deserve to be further explored in the future.}.  As a result, roughly half (in total weight)
of the operators going across the position $y$ will contribute and
hence
\begin{equation}
C(t)=\frac{1}{2} 4  \sum_j  \alpha_j^2(t) \delta( \text{right most site} \ge y )  = 2\int_y^L ds f_R(s,t)
\end{equation}
We therefore show that $\overline{d^2(t)}$ is essentially the
rescaled version of $C(t)$ with the same distance $|y-x|$ in these
chaotic systems
\begin{equation}
\overline{d^2(t)} = \frac{1}{|B|} C( t)
\end{equation}
The numerical verification of the collapse is shown in
Fig.~\ref{fig:pow_d2_oto} and Fig.~\ref{fig:Floq_d2_OTO_L_A}.
Notice that $\overline{C_s(t)}$ we compute is averaged over Page ensemble,
\begin{align}
\overline{C_s(t)} = -\overline{\langle\psi| [\hat{O}(x,-t),\hat{O}^{\prime}(y)]^2|\psi\rangle }
\end{align}
which is equivalent to $C(t)$ defined in Eq.\eqref{C_trace} (using
the identity Eq.\eqref{identity_o}). The collapse is an evidence
that both $\overline{d^2(t)}$ and $C(t)$ measure the area of the
wave front in the chaotic many-body systems.
%%%%%%%%%%%%%%%%%%%%%%
%%%%%%%%%%%%%%%%%%

One interesting question is: how does $\overline{d^2(t)}$ or
$C(t)$ grow with the time? Based on the arguments above, we can
see that it depends on the shape of the operator wave front and
the way it propagates, which is ultimately determined by the form
of the interaction. For spin-$\frac{1}{2}$ chain with local
interaction, we do not find an obvious exponentially growing
regime in $\overline{d^2(t)}$ and $C(t)$. The absence of Lyapunov regime in spin-$1/2$ chain model is also discussed in Ref.\ \onlinecite{Luitz2017} and is different from the classical chaotic spin chain model where we effectively have $S\to\infty$\cite{Wijn2013}. This is consistent with
recent study in the random tensor network with local unitary time
evolution, where  $f_R(s,t)$ in the coarse grained picture is the
probability distribution generated by a biased random
walk\cite{Nahum2017,Keyserlingk2017}. The distribution $f_R(s,t)$
is a moving Gaussian packet with the width of front broadened
diffusively and there is no clear exponentially growing regime in
$C(t)$.

However, non-local power-law interaction changes the story. As
shown in Fig.~\ref{fig:Floq_pow_d2_oto} and
\ref{fig:Ising_pow_d2_oto}, when the distance between $x$ and $y$
is large, we find a clear exponentially growing regime in both
$\overline{d^2(t)}$ and $C(t)$. This is because the power-law
interaction generates a wider wave packet in $f_R(s,t)$. We expect
that when $|y-x|$ is large, the front has a near-exponential
shape. To verify this point, we investigate $\overline{d^2(t)}$
and $C(t)$ for different distances $|y-x|$ at the same time $t$.
As shown in Fig.~\ref{fig:Floq_wavefront}, the area of the wave
front in regime $A$ decreases exponentially as we increase the
distance $|y-x|$. The slope seems to be invariant as we increase
time $t$.

%%%%%%%%%%%%%%%%%%%%%%
%%%%%%%%%%%%%%%%%%

The near-exponential shape of the wave front is responsible for
the exponentially growing regime in $\overline{d^2(t)}$ or $C(t)$,
with the growth rate as the so-called quantum Lyapunov exponent,
which is also found in some quantum many-body models in the large
$N$
limit\cite{Shenker2013a,Shenker2013b,Shenker2014,Maldacena2016,Kitaev2014,Kitaev2015,Sachdev1993}
but is absent in the spin-$\frac{1}{2}$ chain with local
interaction where $N=2$. It seems that the long-range power-law
interaction effectively increases $N$ and gives rise to this
Lyapunov regime. In the future, it would be interesting to have a
better understanding on this Lyapunov regime and explore the
dependence of Lyapunov exponent on the distance $|y-x|$ in some
random non-local unitary tensor networks.

In most of the results, we use the Page states to compute
$\overline{d^2(t)}$ for both simple numerical implementation and
analytical control. They are however difficult to be generated in
experiments. To resolve this issue, we also consider
$\overline{d^2(t)}$ averaged over thermal states, which in
comparison are more physical and much easier to be realized in the
experiments. We prepare the thermal states by evolving the RPS
states under unitary time evolution. After sufficiently long time
evolution, these states will thermalize and can be used as the
initial state for $\overline{d^2(t)}$. These thermal states share
similar properties with the Page states but are subjected to some
conservation laws determined by the Hamiltonian. In
Fig.~\ref{fig:pow_d2_oto}, we see that the thermal state average
matches well with $\overline{d^2(t)}$ of the Page states for a
large regime of time with both the Floquet evolution and time
independent Hamiltonian. Notably, in spin model with non-local
interactions, the thermal state average $\overline{d^2(t)}$ also
grows exponentially with time, showing the same behavior as Page
ensemble.

\section{Conclusion}
\label{conclusion}

In conclusion, we study Hilbert-Schmidt distance $d^2(t)$ between
reduced density matrices of two many-body wave functions initially
differed by a local perturbation. Under unitary time evolution
governed by a chaotic Hamiltonian, $\overline{d^2(t)}$ grows with
time and eventually saturates to a small constant. In contrast to
the exponential divergence of nearby trajectories in classical
chaos, the growth of $\overline{d^2(t)}$ depends on the form of
the interactions in the non-integrable models. We find that for
spin-$\frac{1}{2}$ chain model with non-local power-law
interaction, $\overline{d^2(t)}$ has a clear exponential growing
regime if the local perturbation is far away from the subsystem.
While for spin-$\frac{1}{2}$ chain with local interaction,
$\overline{d^2(t)}$ always grows slower than the exponential
function.

Moreover, we show that after sufficiently long time evolution,
$\overline{d^2(t)}$ will approach a small constant, indicating the
thermalization of a small subsystem in a many-body quantum state.

We further use operator scrambling picture to show that
$\overline{d^2(t)}$ is the rescaled area of the wave front of the
evolved local perturbation operator. The same interpretation also
applies to the square of the commutator $C(t)$ under chaotic
evolution. Hence the growth scaling of $\overline{d^2(t)}$ is
determined by the shape of the wave front, which is ultimately
determined by the form of interactions. From the collapse of the
numerical data of $C(t)$ and $\overline{d^2(t)}$, we conclude that
the exponent in the exponential growing region of
$\overline{d^2(t)}$ is the quantum Lyapunov exponent which was
previously observed in some large $N$ models.

\acknowledgements We acknowledge useful discussion with Thomas
Faulkner, David Huse, Yevgeny Bar Lev, Andreas W.W. Ludwig, Shinsei Ryu and Xueda Wen. XC was
supported by a postdoctoral fellowship from the Gordon and Betty
Moore Foundation, under the EPiQS initiative, Grant GBMF4304, at
the Kavli Institute for Theoretical Physics. TZ was supported by the National Science Foundation under grant numbers DMR-1306011. Cenke Xu is supported by the David and Lucile Packard Foundation, and NSF Grant No. DMR-1151208. We acknowledge
support from the Center for Scientific Computing from the CNSI,
MRL: an NSF MRSEC (DMR-1121053).

\appendix

\section{Average over Page ensemble}
\label{app:aver_over_page}

In this section, we compute the expectation values
$\overline{\langle\psi|\hat{O} |\psi\rangle}$ and
$\overline{\langle\psi|\hat{O}_1|\psi\rangle\langle\psi|\hat{O}_2|\psi\rangle}$
over the Page ensemble. For the ease of notation, we will reserve
$n$ to denote the complex dimension of the Hilbert space. When
comparing the result with main text, $n$ should be identified with
$|H|$.

\subsection{Probability Measure}
The random pure state (Page state) has a probability measure that
is uniform in the Hilbert space. Such a uniform measure has ${\rm
U}(n)$ invariance, meaning that the average is invariant under the
${\rm U}(n)$ action, \eg
\begin{equation}
\overline{\langle \psi | \hat{O} |\psi  \rangle } = \overline{\langle \psi | U^{\dagger} \hat{O}  U |\psi  \rangle}, \quad \forall U \in {\rm U}(n)
\end{equation}
Such invariance can be constructed by choosing complex Gaussian
random variables as the coefficients in an orthonormal basis. For
vector $\vec{c} \in \mathbb{C}^n$, whose component $c_i$ are
sampled from standard complex normal distribution, we have the
required invariance,
\begin{equation}
\begin{aligned}
\mathbb{E}( U \vec{c} ) &= U \mathbb{E}( \vec{c} ) = \vec{0} = \mathbb{E}( \vec{c} )\\
\mathbb{E}( U (\vec{c} - \mu) U ( \vec{c} - \mu )^{\dagger} U^{\dagger}  ) &= U \mathbb{E}( \vec{c} \vec{c}^{\dagger}   ) U^{\dagger}   = \I =  \mathbb{E}( \vec{c} \vec{c}^{\dagger}   )
\end{aligned}
\end{equation}
This is how we generate the Page state in the numerical calculation.

Writing $c_j = x_{2j-1} + i x_{2j}$, the $N = 2n$-dimensional real
vector $\vec{x}$ has independent real Gaussian random variable as
its component. With the normalization constraint, the probability
measure is
\begin{equation}
\begin{aligned}
d \mu \propto& \exp( - \frac{1}{2} |\vec{x}|^2 ) \delta( 1 -
|\vec{x}|^2 ) \prod_i dx_i \propto \delta( 1 - |\vec{x}|^2
)\prod_i dx_i
\end{aligned}
\end{equation}
which is the uniform measure on the $2n - 1$ dimensional sphere.
The expectation value of the state $\overline{ \langle \psi|
\hat{O} |\psi \rangle }$ is a function (polynomial) $f(\vec{x})$
of real and imaginary parts of the elements. Averaging over the
random states is given by the integral
\begin{equation}
 \overline{ \langle \psi|  \hat{O} |\psi \rangle } = \int_{S_{N-1} }  d\mu \,  f( \vec{x} )
\end{equation}

\subsection{Moments of the probability measure}

The $\overline{\langle\psi|\hat{O} |\psi\rangle}$ and
$\overline{\langle\psi|\hat{O}_1|\psi\rangle\langle\psi|\hat{O}_2|\psi\rangle}$
can be converted to the average over the coefficients of the Page
state, e.g.
\begin{equation}
\label{o_exp}
\begin{aligned}
\overline{\langle\psi|\hat{O} |\psi\rangle} &= O_{ij} \overline{c_i^* c_j}= O_{ij} \overline{(x_{2i-1} -i x_{2i} ) (x_{2j-1} + i x_{2j} )}\\
\overline{\langle\psi|\hat{O}_1 \psi \rangle \langle \psi | \hat{O}_2 |\psi\rangle} &= O^1_{ij} O^2_{kl} \overline{(x_{2i-1} -i x_{2i} ) (x_{2j-1} + i x_{2j} ) (x_{2k-1} -i x_{2k} ) (x_{2l-1} + i x_{2l} )}
\end{aligned}
\end{equation}
hence we need up to 4th order moments of $x_i$.

By the reflection symmetry, there are only 3 types of moments that
need to be computed,
\begin{equation}
A_1  =  \int d\mu \, x_i^2, \quad \int d\mu\,  x_i ^2 x_j ^2
= \left\lbrace
  \begin{aligned}
    A_2 & \quad i = j \\
    B_2 & \quad i \ne j \\
  \end{aligned} \right.
\end{equation}
The 2nd moment
\begin{equation}
A_1  =  \int d\mu \, x_i^2 = \frac{1}{N} \int d\mu  |\vec{x}|^2 = \frac{1}{N}
\end{equation}
is be obtained by ${\rm O}(N)$ symmetry, which also relates $A_2$
and $B_2$,
\begin{equation}
1 = \int d\mu \, |\vec{x}|^2 \times |\vec{x}|^2  = N A_2 + N( N - 1 ) B_2
\end{equation}
So we only need $A_2$ or $B_2$.

We first compute the normalization constant, which is proportional
to the area of sphere
\begin{equation}
\begin{aligned}
  \int_{\mathbb{R}^N}  \delta( 1 - | \vec{x}| ^2 ) \prod_i  dx_i  &= \int_0^{\infty} r^{N-1} dr \delta( 1 -r^2  ) {\rm S}_{N-1}  \\
&= \frac{1}{2} {\rm S}_{N-1}  \int_0^{\infty} \delta( 1 - r^2 )  r^{ N - 2} dr^2  = \frac{1}{2} {\rm S}_{N-1}
\end{aligned}
\end{equation}
With this we can compute $A_2$ with the remaining ${\rm O}(N-1)$
symmetry,
\begin{equation}
\begin{aligned}
  A_2 & =    \frac{2}{{\rm S}_{N-1}} \int x_1^4 \delta( 1 - |\vec{x}|^2 ) \prod_i dx_i  = \frac{{\rm S }_{N-2} }{{\rm S }_{N-1} } \int_{0}^{\infty} dr^2 \int_{0}^{\infty} dx^2_1 \, \delta( 1 - x_1^2 - r^2 ) r^{ N - 3}  x^3_1 \\
& =  \frac{{\rm S }_{N-2} }{{\rm S }_{N-1} } \int_{0}^{1} dx\,   x^{\frac{3}{2}} ( 1 - x) ^{\frac{N - 3}{2}}  = \frac{{\rm S }_{N-2} }{{\rm S }_{N-1} }\frac{3 \sqrt{\pi}}{4} \frac{\Gamma( \frac{N-1}{2})}{\Gamma(\frac{N}{2}+2 )} \\
& = \frac{3}{4} \frac{\Gamma( \frac{N}{2} ) }{\Gamma(\frac{N}{2}+2)} = \frac{3}{N( N +2 ) } \implies B_2 = \frac{1}{N( N+2 ) }
\end{aligned}
\end{equation}
In fact, the general pure $2k$ moments can be computed similarly
\begin{equation}
\begin{aligned}
A_{2k} &= \frac{2}{{\rm S}_{N-1}} \int x_1^{2k} \delta( 1 -
|\vec{x}|^2 ) \prod_i dx_i = \frac{(2k-1)!!}{(N + 2k -2 ) (N + 2k
- 4 ) \cdots N }
\end{aligned}
\end{equation}
Now we apply these results to the quantity we need to compute in
Eq.\eqref{o_exp}, where $N = 2n$. We have the 2nd moments
\begin{equation}
\overline{c^*_ic_j} = \overline{(x_{2i-1} -i x_{2i} ) (x_{2j-1} + i x_{2j} )} = \delta_{ij} \frac{2}{N} = \frac{1}{n} \delta_{ij}
\end{equation}
The general 4th moments reduces to three types of contractions
\begin{equation}
\begin{aligned}
\overline{c_i^* c_j c_k^* c_l } &= ( \delta_{ij} \delta_{kl} + \delta_{il} \delta_{jk} ) \overline{c_i^* c_i c_k^* c_k }(1 - \delta_{ik} ) +  \delta_{ik} \delta_{jl} \overline{c_i^* c_i^* c_j c_j } ( 1 - \delta_{ij} ) + \delta_{ij} \delta_{jk}\delta_{kl} \overline{c_i^* c_i c_i^* c_i }
\end{aligned}
\end{equation}
where each of them can be represented by $A_2$ and $B_2$
\begin{equation}
\begin{aligned}
 \overline{c_i^* c_i c_k^* c_k }( 1- \delta_{ik} )   &= (1- \delta_{ik} )\overline{( x_{2i-1}^2 + x_{2i}^2)( x_{2k-1}^2 + x_{2k}^2) } \\
 &= ( 1 - \delta_{ik})4 B_2 = ( 1 - \delta_{ik} ) \frac{1}{n(n+1)}  \\
 \overline{c_i^* c^*_i c_j c_j }( 1 - \delta_{ij} )  &=  ( 1 - \delta_{ij} ) \overline{( x_{2i-1}^2 - x_{2i}^2 - 2i x_{2i-1} x_{2i} )( x_{2j-1}^2 - x_{jk}^2 + 2ix_{2j-1} x_{2j} ) } \\
&= ( 1 - \delta_{ij} ) (2B_2 - 2B_2 ) = 0\\
\overline{c_i^* c_i c_i^* c_i } &= \overline{( x_{2i-1}^2 + x_{2i}^2)^2 }  =  \frac{2}{n(n+1)}
\end{aligned}
\end{equation}
Therefore we have "Wick theorem" for the moments of $c_i$
\begin{equation}
\begin{aligned}
  \overline{c_i^* c_j c_k^* c_l } &= ( \delta_{ij} \delta_{kl} + \delta_{il} \delta_{jk} ) \frac{1}{n(n+1)} (1 - \delta_{ik} ) + \frac{2}{n(n+1)} \delta_{ik} \delta_{jl} \delta_{ij} \\
 &=  ( \delta_{ij} \delta_{kl} + \delta_{il} \delta_{jk} ) \frac{1}{n(n+1)}
\end{aligned}
\end{equation}

\subsection{Average of expectation values}

We apply the results of the previous sections to the average of
expectation values. For operator $\hat{O}$, we have
\begin{equation}
\begin{aligned}
\overline{\langle \psi| \hat{O}|\psi \rangle}  &= O^{ij} \overline{  c_i^* c_j }  =  \frac{1}{n} \delta_{ij} O^{ij} =  \frac{1}{n}\Tr( \hat{O} )
\label{identity_o}
\end{aligned}
\end{equation}
For two operators $\hat{O}_1$ and $\hat{O}_2$, we have
\begin{equation}
\begin{aligned}
\overline{\langle \psi| \hat{O}_1 |\psi \rangle \langle \psi| \hat{O}_2 |\psi \rangle} &= O_1^{ij} O_2^{kl} \overline{c_i^* c_j c_k^* c_l } = \frac{1}{n(n+1)} O_1^{ij} O_2^{kl}(\delta_{ij} \delta_{kl} + \delta_{il}\delta_{jk}) \\
&=  \frac{1}{n(n+1)}[\Tr( \hat{O}_1 \hat{O}_2  ) + \Tr(\hat{O}_1) \Tr(\hat{O}_2 )]
\label{identity_oo}
\end{aligned}
\end{equation}
where $n$ is the Hilbert space dimension $|H|$ in the main text.

\section{Averaged distance of two independent Page states}
\label{app:aver_dist}

As discussed in App.~\ref{app:aver_over_page}, the random pure
state (Page state) can be constructed by assigning complex
Gaussian random variables as the coefficients in any orthonormal
basis.

We divide the Hilbert space into subsystem $A$ with dimension $n$
and $B$ with dimension $m \ge n$. In the same orthonormal basis,
we can obtain the decomposition coefficients $\psi_{ij}$ of the
wavefunction on the subsystem basis. Call this $n \times m $
matrix $Y$, the reduced density matrix is $\hat\rho_A = Y
Y^{\dagger}$. The reduced density matrix consists of products of
Gaussian random variables, and thus is a Wishart matrix with unit
trace constraint.

The element of $Y$ matrix satisfies a Gaussian probability
distribution\cite{nadal_statistical_2011}
\begin{align}
P(\{Y_{ij}\})\sim \exp\left[\frac{\beta}{2}\Tr YY^{\dag} \right]
\end{align}
where the Dyson index $\beta = 2$ in this case. Therefore the
eigenvalue of $YY^{\dag}$ scales as $n$, and the (rescaled)
spectral density is defined as
\begin{equation}
\varrho(x, \rho ) = \frac{1}{n} \sum_i \delta( x - \frac{\lambda_i}{n} ).
\end{equation}
In the large $n$ limit (keeping the ratio $\alpha = \frac{n}{m}$ fixed), the spectral density reduces to the Marchenko-Pastur distribution with parameters
\begin{equation}
\varrho_{\,\!_{\rm W}}( x ) = \frac{1}{2\pi x} \sqrt{ (\frac{ \alpha_+}{\alpha} -x) ( x - \frac{\alpha_{-}}{\alpha }  )}
\end{equation}
where $\alpha_{\pm} = (1 \pm \sqrt{\alpha} )^2 $.

Wishart matrix with the trace constraint is called fixed trace Wishart-Laguerre ensemble. Due to the constraint, the eigenvalues scales as $\frac{1}{n}$, so we shall instead use\cite{nadal_statistical_2011,mejia_difference_2017}
\begin{equation}
\varrho(x, \rho ) = \frac{1}{n} \sum_i \delta( x - n  \lambda_i  )
\end{equation}
The corresponding asymptotic distribution is
\begin{equation}
  \varrho_{\,\!_{\rm W, tr}}( x ) = \frac{1}{2 \pi \alpha x } \sqrt{( \alpha_{+} - x ) (  x- \alpha_{-} ) }
\end{equation}
it is different from the one without the trace constraint.

The distance we want to compute is the second moments of the
eigenvalues of $\hat\rho_1 - \hat\rho_2$
\begin{equation}
d^2(t) = \Tr ( \hat\rho_1 - \hat\rho_2)^2  =  \sum_i \lambda_i^2
\end{equation}
which is the distribution of the {\it difference} of two independent variables sampled from the Marchenko-Pastur distribution. Because of the constraint $\sum_i \lambda_i = 0$, the eigenvalues also scale as $\frac{1}{n}$,
\begin{equation}
\varrho(x, \hat\rho_1 - \hat\rho_2  ) = \frac{1}{n} \sum_i \delta( x - n  \lambda_i  )
\end{equation}
Taking a function $f(x)$, the average of the sum $\sum_i f(\lambda_i)$ can be carried over the spectral density function
\begin{equation}
\begin{aligned}
  \frac{1}{n} \sum_{i=1}^n f(\lambda_i )  &=  \int \, f( \lambda ) \frac{1}{n} \sum_{i=1}^n \delta( \lambda - \lambda_i ) P( \{\lambda_i \} )\,  d\lambda \\
&= \int \, f( \lambda ) \frac{1}{n} \sum_{i=1}^n \delta( n\lambda - n\lambda_i ) P( \{\lambda_i \} )\,  d n \lambda \\
&= \int   f( \frac{x}{n} ) \varrho( x, \hat\rho_2 - \hat\rho_1 ) \, dx
\end{aligned}
\end{equation}
Ref.~\onlinecite{mejia_difference_2017} computes the asymptotic
eigenvalue distribution and its absolute moments
\begin{equation}
m_z = \int |x|^z \varrho( x ) \, dx
= \frac{\Gamma( z + 1) ( 2\alpha )^{\frac{z}{2}} }{\Gamma( \frac{z}{2} + 1 ) \Gamma (\frac{z}{2} + 2 )} {}_2 F_1 ( 1 - \frac{z}{2}, -\frac{z}{2} ;  \frac{z}{2} +2 ; \frac{\alpha}{2} )
\end{equation}
where ${}_2 F_1$ is the hypergeometric function. Hence
\begin{equation}
\begin{aligned}
  d^2 &= n  \int \frac{1}{n^2} x^2  \varrho(x) \, dx  = \frac{1}{n} m_2 = \frac{1}{n}(2\alpha) {}_2 F_1 (0, -1; 3; \frac{\alpha}{2})\\
&= \frac{2 \alpha }{n}  = \frac{2}{m}
\end{aligned}
\end{equation}

\bibliographystyle{unsrt}
\bibliography{d_squ}

\end{document}